\documentclass[10pt,journal,compsoc]{IEEEtran}
\ifCLASSOPTIONcompsoc
  \usepackage[nocompress]{cite}
\else
  \usepackage{cite}
\fi
\ifCLASSINFOpdf
\else
\fi
\usepackage{amsmath}
\usepackage{graphicx}
\usepackage{hhline}
\usepackage{subcaption}

\begin{document}
%
\title{Efficient State-space Exploration in Massively Parallel Simulation Based Inference}
%
%
%

\author{Sourabh~Kulkarni,~\IEEEmembership{Member,~IEEE,}
        and~Csaba~Andras~Moritz,~\IEEEmembership{Senior Member,~IEEE}
\IEEEcompsocitemizethanks{\IEEEcompsocthanksitem S. Kulkarni and C.A. Moritz are with the Department
of Electrical and Computer Engineering, University of Massachusetts, Amherst, MA.\protect\\
E-mail: skulkarni@umass.edu
}
\thanks{Manuscript received April 19, 2005; revised August 26, 2015.}}

%
%

\markboth{Transactions on Parallel and Distributed Systems,~Vol.~xx, No.~x, Month~2021}%
{Shell \MakeLowercase{\textit{et al.}}: Bare Demo of IEEEtran.cls for Computer Society Journals}
%



\IEEEtitleabstractindextext{%
\begin{abstract}
Simulation-based Inference (SBI) is a widely used set of algorithms to learn the parameters of complex scientific simulation models. While primarily run on CPUs in HPC clusters, these algorithms have been shown to scale in performance when developed to be run on massively parallel architectures such as GPUs. While parallelizing existing SBI algorithms provides us with performance gains, this might not be the most efficient way to utilize the achieved parallelism. This work proposes a new algorithm, that builds on an existing SBI method - Approximate Bayesian Computation with Sequential Monte Carlo(ABC-SMC). This new algorithm is designed to utilize the parallelism not only for performance gain, but also toward qualitative benefits in the learnt parameters. The key idea is to replace the notion of a single 'step-size' hyperparameter, which governs how the state space of parameters is explored during learning, with step-sizes sampled from a tuned Beta distribution. This allows this new ABC-SMC algorithm to more efficiently explore the state-space of the parameters being learnt. We test the effectiveness of the proposed algorithm to learn parameters for an epidemiology model running on a Tesla T4 GPU. Compared to the parallelized state-of-the-art SBI algorithm, we get similar quality results in $\sim 100$x fewer simulations and observe $\sim 80$x lower run-to-run variance across 10 independent trials. 
\end{abstract}

\begin{IEEEkeywords}
Parallel Algorithms, Approximate Bayesian Computation, Simulation-based Inference, Likelihood-free Inference, Bayesian Inference, Statistical Machine Learning, Stochastic Optimization, Epidemiology, Compartmental Models, COVID-19, High Performance Computing, GPU.
\end{IEEEkeywords}}

\maketitle

\IEEEdisplaynontitleabstractindextext

%
\IEEEpeerreviewmaketitle

\IEEEraisesectionheading{\section{Introduction}\label{sec:introduction}}

\IEEEPARstart{S}{imulation}-Based Inference (SBI) shows great promise in a wide variety of scientific domains. It is used to learn the parameters of complex simulation models which attempt to capture the underlying physical processes. The learned parameters can then be used to generate predictions from the model. To give a few examples, SBI can be used to learn key parameters for elementary particle collision experiments \cite{CMS_CERN}, in biology to study the dynamics of protein networks\cite{abc_biochem}, and  in epidemiology for learning the key parameters of an epidemic such as COIVD-19 to understand and predict the spread as well as to inform large-scale interventions \cite{Warne2020}. Other areas of science where SBI algorithms are used are cosmology \cite{abc_cosmology,Galactic_simulations}, cognitive science \cite{abc_cognitive}, econometrics \cite{abc_econometrics}, and systems biology \cite{abc_biochem,abc_evolution,abc_evolution2,abc_sysbio}.

SBI algorithms have been widely used in these disciplines because, unlike other statistical inference algorithms, SBI does not require additional criteria to be satisfied to perform parameter learning for a model – simply the ability to simulate the model suffices. Typically, alternatives to SBI algorithms are the family of Markov Chain Monte Carlo algorithms (MCMC) \cite{hamra2013markov}, which require the ability to compute the ‘likelihood function’ over the model, which is intractable for many scientific models \cite{sim_infer_survey}. For example, in modelling an epidemic, we must consider subpopulations which are often not observable (e.g., the number of untested infections), and in doing so, the epidemiological model ends up having an intractable likelihood function. In such scenarios, SBI algorithms are the only option for parameter learning. 

Primarily, these SBI methods are implemented on CPUs \cite{Warne2020_github}. This is not efficient, as these algorithms have a lot of potential for parallelism and hence can benefit from GPUs and other massively parallelized compute architectures. At the core of all simulation-based inference algorithms is the ability to compute a large number of simulations with randomly sampled parameters. The simulations are then compared with real-world observations to gauge the quality of the parameters used in the simulation. The algorithms then iteratively converge to the best possible parameters. These massive number of simulations can, in most SBI algorithms, be performed in an embarrassingly parallel fashion per iteration \cite{sim_infer_survey}.  

Parallelized versions of existing SBI algorithms have been developed and have shown to provide performance benefits \cite{OG_ABC_IPU, kulkarni2020hardwareaccelerated}. However, they are not the best ways to utilize the achieved parallelism, as they were originally designed to perform best in the regime of progressing one simulation at a time. As part of this work, we perform experiments with these massively parallel versions of existing SBI algorithms to show that they have inconsistent results across independent runs. They also tend to get stuck in local minima and hence provide sub-optimal results. With the ability to perform up to 100K simulations in parallel at a time \cite{OG_ABC_IPU, kulkarni2020hardwareaccelerated}, we require novel ways to obtain the best possible progress from them and to avoid the issues that arise with using parallelized versions of existing algorithms.  

In this work, we propose a new simulation-based inference algorithm, which is designed to perform best in such a massively parallel regime. The algorithm builds on an existing SBI algorithm, known as Approximate Bayesian Computation with Sequential Monte Carlo (ABC-SMC). The proposed algorithm exploits the achieved parallelism not only for performance gain, but also for a qualitative improvement in the results achieved. The key idea of this new algorithm is an efficient method of parameters’ state-space exploration; this is achieved by a novel way to repurpose the notion of ‘step-size’ in massively parallel stochastic optimization processes. 

 The ‘step-size’ (also known as the ‘learning rate’ in some literature), in SBI and other stochastic optimization algorithms in general, is an important hyperparameter. This hyperparameter, at each stage of the algorithm, captures the amplitude of the change introduced to the parameters being evaluated. In current methods, the step size is usually a single number, and is tuned carefully as the algorithm progresses, usually having greater values initially, and lower values as the algorithm approaches convergence. The key contribution of this work is the introduction of a new concept – step–size distribution. When 100K simulations occur in parallel, using a single step size does not introduce enough ‘diversity’ in the parameters to make efficient use of all those simulations toward the convergence process. The proposed method, on the other hand, allows each of those 100K simulations to have an independent step-size which is sampled from the step-size distribution, which is a tuned Beta distribution. Instead of using complex methods for tuning the single step size, the proposed method uses simple ratios that encode the progress of the algorithm to tune the PDF of the Beta step-size distribution. We call this new algorithm 'Parallel ABC-SMC with Beta-Distributed Step-Sizes', or P-ABC-SMC BDSS in short.

P-ABC-SMC BDSS is evaluated for its effectiveness to perform parameter learning for a stochastic epidemiology model used for understanding and predicting the spread of COVID-19. The algorithm is compared with the parallelized version of the current state-of-the-art ABC-SMC algorithm, where step-size tuning is performed using MCMC. This baseline is called P-ABC-SMC MCMC in short. Experiments are performed on a Nvidia Tesla T4 GPU. We show that the new method provides better quality of learned parameters in fewer number of simulation steps, and better consistency across independent trials. Experiments of 10 independent trials showed consistently better results - the P-ABC-SMC BDSS obtained better results than P-ABC-SMC MCMC in $\sim 100$x fewer simulations, and $\sim 80$x lower run-to-run variance. Experiments also confirm that P-ABC-SMC BDSS’s performance efficiency improves with increasing degree of parallelism. 

While we use the epidemiology model as an example, P-ABC-SMC BDSS should be effective wherever the ABC-SMC method is used. Furthermore, this work shows promise not only for the ABC-SMC algorithm, but for all stochastic optimization algorithms. In these algorithms, there exists a method of tuning step-sizes. These methods could be replaced instead by the notion of a tuned step-size distribution - like the one presented in this work. This could open up new possibilities of developing parallelized versions of these stochastic optimization algorithms - consequently obtaining performance gains on contemporary parallel hardware architectures.

The rest of the paper is organized as follows: Section II provides background information on the epidemiology model used to test the new algorithm, and the basics of SBI algorithms and the current state-of-the-art ABC-SMC MCMC, over which the proposed algorithm improves upon; Section III Describes in detail the proposed P-ABC-SMC BDSS algorithm and its implementation; Section IV Describes the experimental setup, and the results obtained; Section V discusses the results and implications of this work to the wider domain of stochastic optimization in the massively parallel regime.  

\section{Background}

In this work, we are proposing a new algorithm to perform SBI on a massively parallel scale. We are demonstrating the effectiveness of this algorithm by comparing it to a parallelized version of the current state-of-the-art SBI. To make this comparison, both are used in an epidemiological application to understand and predict the spread of COVID-19 in a nation given their case data. This is done by using a stochastic epidemiological model, which can be used to simulate, on a per day basis, the case outcomes of a population given that we know the key epidemiological parameters. Hence, the task for both the SBI algorithms is to find a distribution over parameters that can i) best explain the currently observed case data for a certain country, and ii) accurately predict the future case data for that country.  

In this section, we shall first establish the mathematical notations that will be used throughout the paper, which are typical of these statistical inference methods, but differ from other ML literature. Secondly, we describe in detail the epidemiological model. Then we shall discuss how SBI algorithms are used to learn parameters for the model using real-world case data. Third, we shall discuss the current SBI algorithms – the Approximate Bayesian Computation (ABC) and a more advanced version – the ABC with Sequential Monte Carlo (ABC-SMC).  

\subsection{Mathematical Notations for Statistical Inference} 

\begin{figure*}[htbp!]
\centerline{\includegraphics[width=\linewidth]{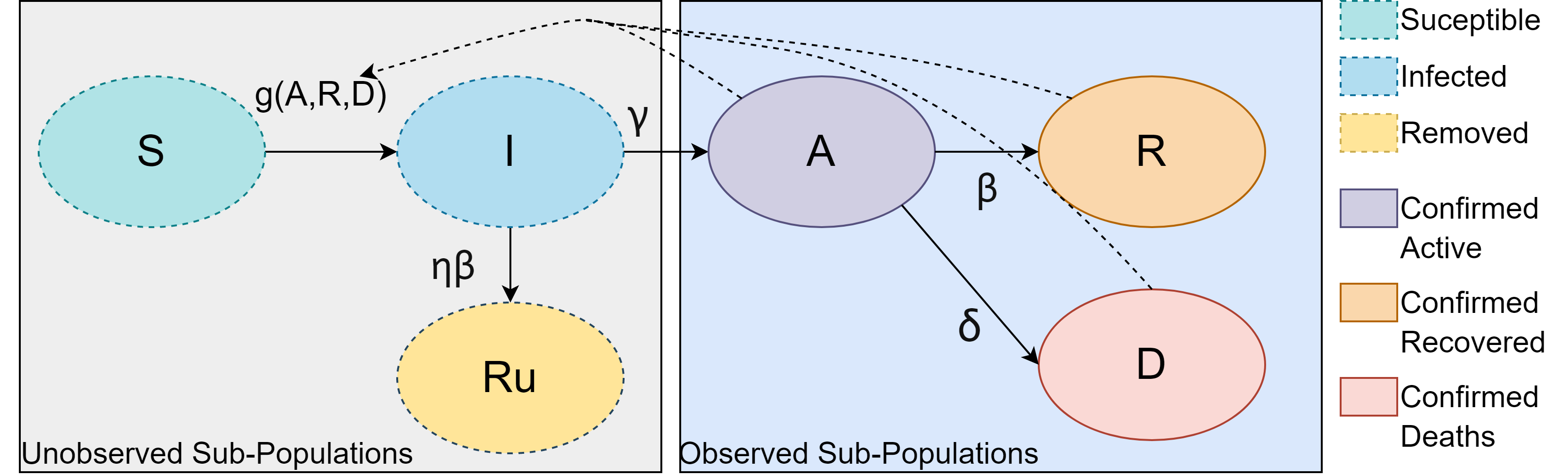}}
\caption{Overview of the epidemiology model flow. The population is divided into $6$ sub-populations. 
On a daily basis, the number of people moving from one sub-population to the other are simulated with a Poisson process, where the rate parameters of these processes are given by the current sub-populations and the transmission parameters (infection rate, death rate, etc.). 
The transition from susceptible to infected is a function of the observed sub-populations 
which captures the response of a population to an increasing number of cases \cite{kulkarni2020hardwareaccelerated}. 
}
\label{f:model_overview}
\end{figure*}

In statistical inference, a model is defined to be a joint probability distribution over its parameters (denoted by $\theta$), and its observed variables (denoted by $x$). The model as a probability distribution is hence denoted as $p(\theta,x)$. The parameters are known to exist in a certain domain a priori, either due to expert knowledge, or according to how the model is designed. This distribution in which the parameters belong before the learning process starts is known as the prior and is denoted by $\pi(\theta)$. The models are generative in nature and can be simulated for a given set of parameters to generate observations $D_s \sim p(x|\theta)$. The evidence, or real-world observations, are denoted by $D_o$. The likelihood of observing $D_o$ for a given set of parameters theta is called the likelihood function $L = p(D_o|\theta)$. The distribution obtained after the parameters are learnt using the observed data $D_o$ is called the posterior $p(\theta|x=D_o)$. 

It is important to note that these statistical `inference' algorithms are solving a `learning' problem, and the expression `parameter inference' in the statistics literature corresponds to `parameter learning' in ML literature. This should not be confused with `inference' as used in the ML literature, which often means the process of estimating an output with a model, given the input. In this paper, the term 'parameter inference' from statistical inference literature is referred to as 'parameter learning'.

\subsection{Stochastic Epidemiology Model for COVID-19}
\label{s:epi}

The epidemiology model considered in this work belongs to the class known as compartmental models. In this model class, the population is divided into several subpopulations, and the spread of infectious disease is modelled as the flow of people from one sub-population to other. 

The model considered in this work~\cite{Warne2020} attempts to capture the spread of COVID-19 
using six subpopulations, three observed and three unobserved. We include a brief overview of the model, but refer the reader to the Supplementary Material of \cite{Warne2020} for further detail.
The transmission across these sub-populations is modeled with a Poisson process approximated in discrete 1-day timebins using the tau-leaping method \cite{tau_leaping}. 
The model consists of $8$ parameters:

\begin{equation}
    \theta = [\alpha_0, \alpha, n, \beta, \gamma, \delta, \eta, \kappa]
\label{parameters}
\end{equation}

with a uniform prior distribution:

\begin{equation}
    \pi = p(\theta) = U(0, [1, 100, 2, 1, 1, 1, 1, 2])
\end{equation}

These prior values were taken as-is from the original model description~\cite{Warne2020}.
They signify the reasonable ranges in which the parameters of interest could lie. 
When we simulate the model with a sample set of these parameters,we get the following state vector:

\begin{equation}
X = [S, I, A, R, D, R^u],
\end{equation}
which consists of the sub-populations of
\textbf{S}usceptible people, undocumented \textbf{I}nfected, 
\textbf{A}ctive confirmed cases, confirmed \textbf{R}ecoveries,
confirmed fatalities \textbf{D}ying,
and unconfirmed \textbf{R}$^u$emoved. The Removed sub-population, $R^u$, comprises those who have recovered or died, but have not been tested. This simulation is typically performed over several days or months and the generated data can be compared with the real-world values of the observable subpopulations.

One of the key challenges in this model is that the state vector $X$ is \textit{partially} observed; i.e. only the $A, R, D$ values are available from observed data.
This makes the likelihood function $p(D|\theta)$ intractable for this model, as the unobserved sub-populations of the model $S, I, R^u$ are required to be `integrated-out' - the likelihood function is defined over \textit{all possible values} of those three subpopulations, and the only way to obtain the likelihood values for use in the learning process would be to integrate the likelihood function over all possible values of those three sub-populations.
Instead, simulation-based inference such as ABC is used to perform inference over this model (See next subsections).

Parameter $\alpha_0$ refers to the base infection rate, while $\alpha$ is the coefficient of the function that captures the changes in infection rate based on the observed sub-populations (A,R,D). $n$ is the exponent t
o the function. Based on these parameters, the total infection rate is assumed to follow:

\begin{equation}
    g_{(A,R,D)} = \alpha_0 + \frac{\alpha}{1 + (A+R+D)^n}
\label{total_infection_rate}
\end{equation}

This function can be modified to capture additional changes to the infection rate based on $A$, $R$, $D_o$ values or even using external data.

The parameters $\gamma$, $\beta$, and $\delta$ are the positive test rate, recovery rate and fatality rate respectively.  
The parameter $\eta$ captures the effectiveness of testing protocols, 
as the rate for unconfirmed infected to be recovered without ever being confirmed is given by $\eta\beta$. 
The initial value parameter, $\kappa$, encodes the number of unobserved infected cases, as a fraction of $A$, at the start of the simulation.

The underlying COVID-19 time-series data, 
provided by Johns Hopkins University~\cite{Dong2020},
contains daily case numbers for $[A, R, D]$.

In its first step, the model initializes the remaining variables
with $R^u=0$, $I_0 = \kappa * A_0$, and 
$S=P-(A_0+R_0+D_0+I_0)$ with $P$ being the total population count at
the first time point.

The second step is to calculate the hazard function $h$
which provides the average update of the model parameters
within one day

\begin{equation}
    h(S, I, A, R, D, R^u)=
    \left(g S\frac{I}{P},
    \gamma  I,
    \beta  A,
    \delta  A,
    \beta \eta I
    \right).
\end{equation}
with $g$ described in equation \ref{total_infection_rate}.

The third step is to sample the transmission numbers from a Poisson distribution with the rate parameter set according to these average numbers.
Instead of a Poisson sampling with $h$ as parameter,
we chose an approximation with 
normal distributions with mean $h$ and variance $\sqrt{h}$
and use the floor of the numbers. This approximation allows us to perform highly optimized parallel simulations as seen in the next section.

The fourth step is to apply the sampled transmission amounts to obtain updated
numbers for the next day 
($S\rightarrow I$, $I\rightarrow A$, $A\rightarrow R$, $A\rightarrow D$,
$I\rightarrow R^u$, ordering according to $h$ function).

The second to fourth steps are repeated for the number of days for which case data is available. 
Eventually, the numbers for $A$, $R$, and $D_o$ can be compared to the real
measurements.

\subsection{Approximate Bayesian Computation}
\label{s:ABC}
The Bayesian statistical inference approach of learning parameters is through obtaining the \textit{posterior} over parameters $\theta$ for a model $p(\theta, x)$ and given observations $D_o$, which is given by Bayes' rule,

\begin{equation}
    p(\theta|D_o) = \frac{p(D_o|\theta)p(\theta)}{p(D_o)} 
\end{equation}
As discussed earlier, the likelihood function $p(D_o|\theta)$ is intractable, since $S_t$, $I_t$, and $R^u_t$ are unobserved. This precludes some approximate Bayesian inference methods such as MCMC.
In the ABC approach, the model simulations are utilized to perform parameter learning. 
First, we sample the parameters $\theta$ from their prior $\theta^* \sim \pi(\theta)$.
Next, we simulate a forward pass of the simulator (as described in Section~\ref{s:epi} for example) to generate observations $D_s \sim p(x|\theta=\theta^*)$ over the number of days we have data for. 
The simulated observations are then compared to the real-world evidence using a distance function $dist(D_s, D_o)$. 
For this model we used the Euclidean distance~\cite{Warne2020}. 
Finally, the sampled parameters $\theta^*$ are accepted 
if the distance function is less than a certain tolerance value 
$\epsilon$, $dist(D_s, D_o) \leq \epsilon$. 
This is repeated until we accept the target number of posterior samples. 
It can be shown that as tolerance $\epsilon$ approaches $0$, the approximate posterior converges to the true posterior~\cite{ABC_paper}. The approximation is also better with more number of posterior samples.

\begin{figure}[htbp!]
\centering
\includegraphics[width=175pt]{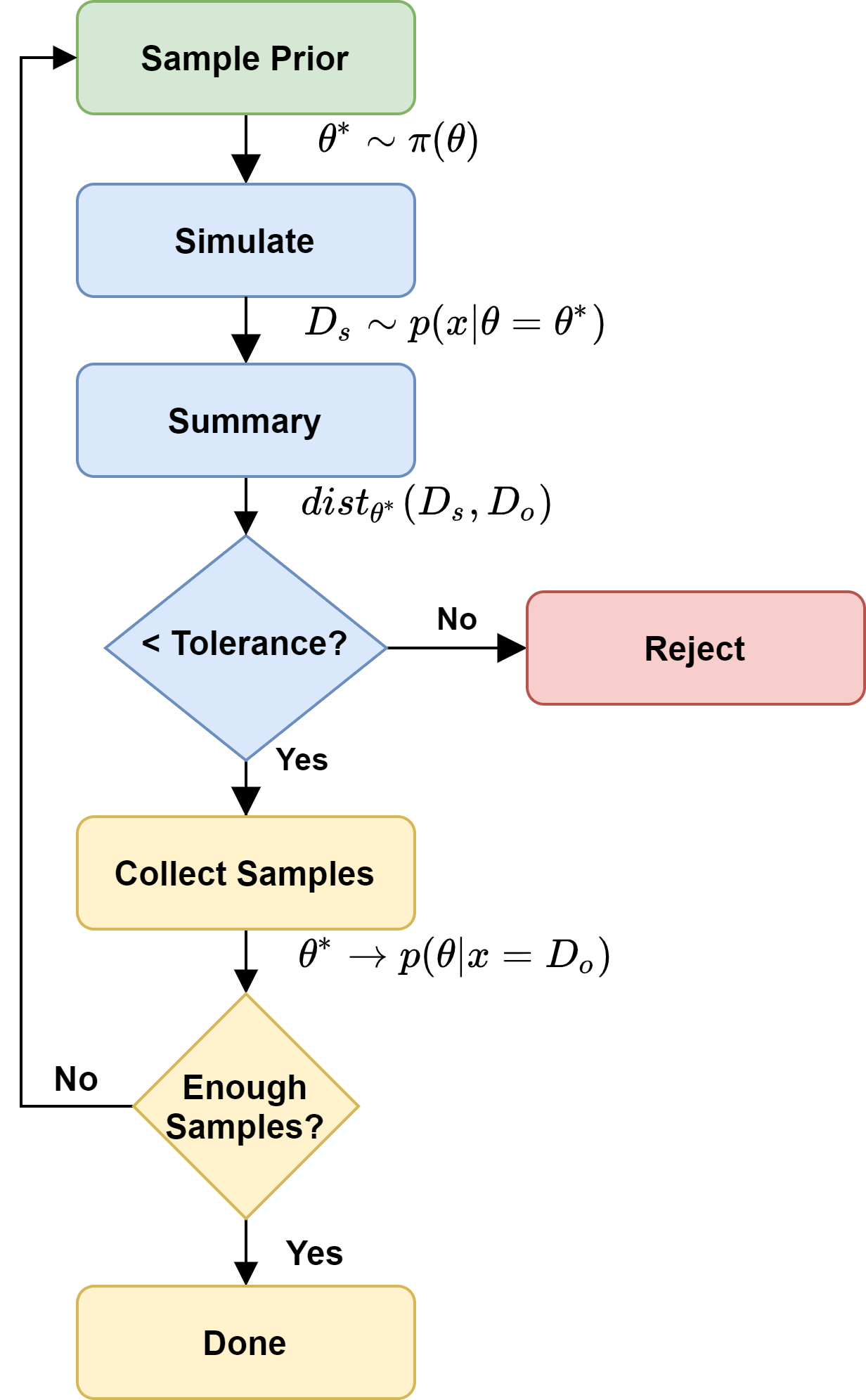}
\caption{Overview of the Approximate Bayesian Computation (ABC) algorithm.}
\label{f:ABC_overview}
\end{figure}

Fig. \ref{f:ABC_overview} shows an overview of the ABC process as a flow chart.
To summarize, in ABC we aim to obtain samples from an approximation to the posterior:
\begin{equation}
    p(\theta|x=D_o) \approx p(\theta|\text{dist}(D_o,D_s)\leq\epsilon) 
\end{equation}
where $D_o$ is the ground truth data, $D_s$ is simulated data depending on $\theta$, and $p(\theta)$ is the prior~\cite{Warne2020}.
The $\text{dist}$ function is the Euclidean distance~\cite{Warne2020}.

The number of simulations that need to be performed to obtain samples below a certain tolerance value increases exponentially, the tolerance value goes lower. Hence using vanilla ABC algorithm becomes infeasible to achieve arbitrarily small tolerance values.  
Instead of choosing a fixed tolerance, sequential Monte Carlo can be 
used to transform an initial set of samples
to a high quality set with a decreasing sequence of tolerances $\epsilon$
and using ABC. This algorithm is called SMC-ABC~\cite{Warne2020, Drovandi2011}, and shall be explained in the next subsection.

\subsection{ABC with Sequential Monte Carlo}

By combining the ABC algorithm with the Sequential Monte Carlo process, we get ABC-SMC. In essence, the ABC-SMC algorithm performs repeated applications of the Bayes rule (see Eq. 6) via the ABC process to obtain better and better quality posterior samples. In ABC-SMC, the algorithm starts with a large tolerance value $\epsilon^0$. Then, through a sequence of several stages, the algorithm obtains sequentially lower tolerance values until the target tolerance value $\epsilon^t$ is reached. In each of its stages $i$, ABC-SMC processes the posterior of previous stage $p^{i-1}(\theta|x=D_o)$ to form the prior of the current stage $\pi^i(\theta)$. In turn, the posterior of current stage becomes the prior of the next stage.

\begin{figure}[htbp!]
\centering
\includegraphics[width=250pt]{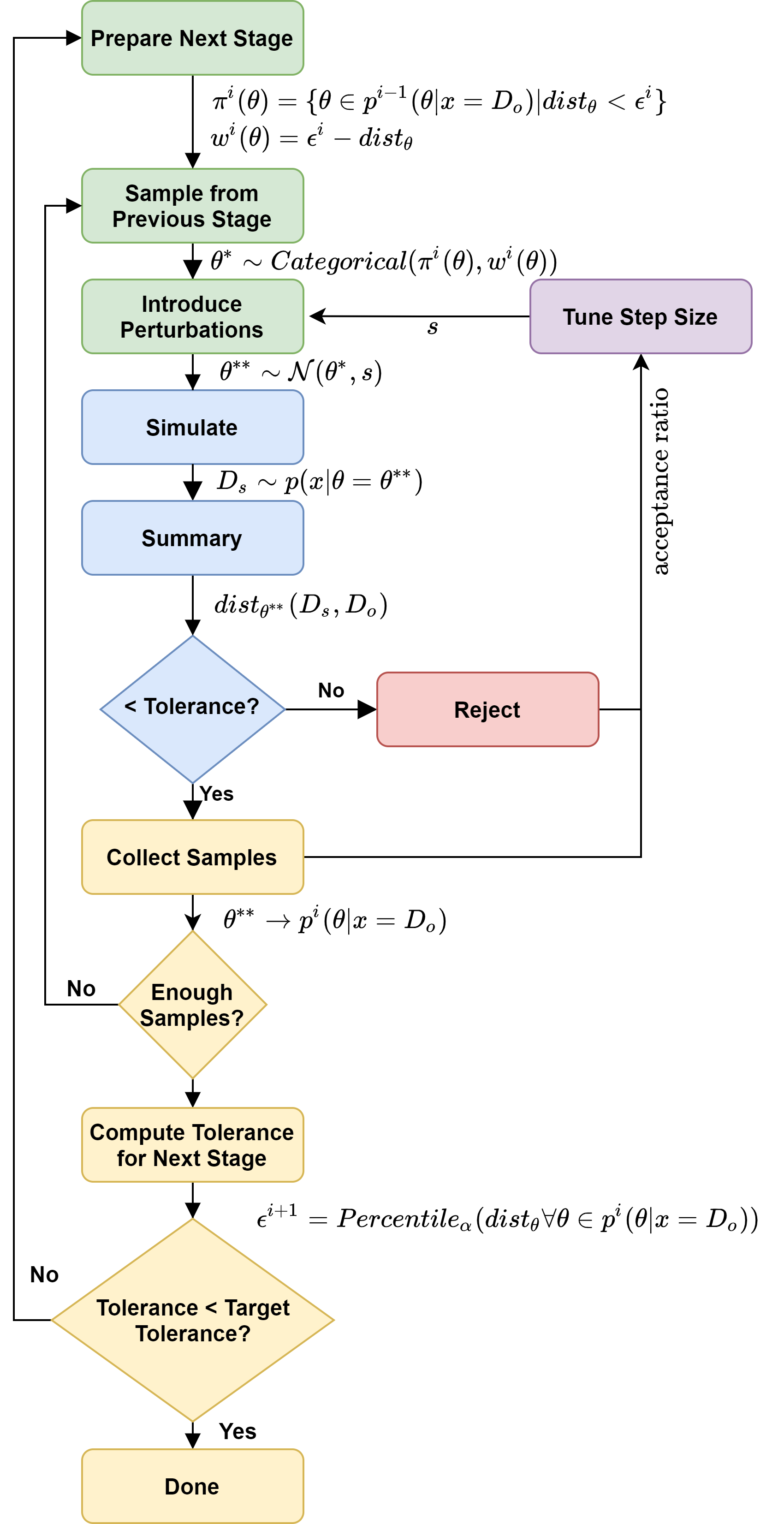}
\caption{Overview of the Approximate Bayesian Computation with Sequential Monte Carlo (ABC-SMC) algorithm.}
\label{f:ABC_SMC_Overview}
\end{figure}

Fig. \ref{f:ABC_SMC_Overview} provides an overview of the ABC-SMC algorithm. As the figure shows, the algorithm builds on the primary ABC framework, and each stage of the ABC-SMC resembles the ABC algorithm. Key differences between ABC and an ABC-SMC stage are: i) preparation of current stage's prior from previous stage's posterior, ii) how parameters are sampled from this prior, iii) the notion of introducing perturbations to sampled parameters, and iv) how the tolerance is computed for the next stage based on the samples accepted in the current stage.
We shall now go through in detail how each of these steps are performed.

The initial stage of ABC-SMC, since there is no previous stage to obtain samples from, is a simple ABC stage with a large tolerance value $\epsilon^0$. When the required number of samples $N$ are obtained, they are used to determine the tolerance for next stage $\epsilon^1$. For all stages, (including the $0^{th}$ stage) This is done with the help of the survival ratio $\alpha$:
\begin{equation}
    \epsilon^{i+1} =  Percentile_{\alpha} (dist_{\theta} \forall \theta \in p^i(\theta|x=D_o))
\end{equation}
Here, the survival ratio $\alpha$ is a hyperparameter. For the purposes of this work, we set it to $\alpha = 0.5$. Hence, at each stage, the distance metric $dist(D_s,D_o)$ of the $50^{th}$ percentile (i.e. median) of the accepted samples at stage $i$ becomes the tolerance for next stage $\epsilon^{i+1}$. Naturally, for all values of $\alpha \in (0,1)$ the tolerance values are strictly non-increasing.

At the start of all stages other than $0^{th}$, the prior of current stage $i$ is computed using the posterior of previous stage $p^{i-1}(\theta|x=D_o)$ and the tolerance of current stage $\epsilon^{i}$ as follows:
\begin{equation}
    \pi^i(\theta) = \{ \theta \in p^{i-1}(\theta|x=D_o)| dist_\theta < \epsilon^{i}\}
\end{equation}
i.e., the prior of current stage $\pi^i(\theta)$ is the set of all parameters $\theta$ in the posterior of previous stage $p^{i-1}(\theta|x=D_o)$, that have a distance metric $D_{\theta}$ smaller than the tolerance $\epsilon^i$. For a survival ratio $\alpha = 0.5$, only the top half of the samples from previous stage's posterior are included in the current stage's prior. 

Next, each of the samples in the current stage's prior $\pi^i(\theta)$ are assigned a weight $w^i(\theta)$:
\begin{equation}
    w^i(\theta) = \epsilon^{i} - dist_\theta
\end{equation}
Hence, a sample is given a higher weight, the lower its distance metric is compared to the current tolerance $\epsilon^{i}$.
These weights are then used to perform weighted random draws $\theta^*$ from $\pi^i(\theta)$:
\begin{equation}
    \theta^* \sim Categorical(\pi^i(\theta), w^i(\theta))
\end{equation}
After drawing a sample from prior $\theta^*$, a perturbation is introduced using a Gaussian random walk:
\begin{equation}
    \theta^{**} \sim \mathcal{N}(\theta^*, s)
\end{equation}
Here, the variance of the Gaussian used to perform a random walk step is called the step size $s$. This ensures that new samples are explored in the vicinity of the samples accepted in the last stage in hopes of finding samples with a lower distance metric. For the perturbations to be effective, the step size needs to be tuned carefully - too small, and the ABC-SMC process gets stuck in a local minima; too large, and the process shall no longer obtain any good samples.

After obtaining a perturbed sample $\theta^{**}$, the subsequent steps are similar to ABC - simulate model, compute distance function, and perform accept/reject. We also maintain the ratio of accepted to rejected samples $\theta^{**}$, which is used to tune the step size $s$. It is here, at the steps of tuning the step size and introducing perturbations, where the proposed algorithm differs from the current state-of-the-art. Hence, to understand the significance of the proposed algorithm, it is essential we review in-depth the step-size tuning and perturbation steps as they are in the current state-of-the-art algorithm.

\subsubsection{MCMC Step Size Tuning and Perturbations}
In the state-of-the-art version of ABC-SMC \cite{Warne2020}, the step size tuning is performed by using Metropolis-Hastings (MH) Algorithm \cite{MHAlgo}, which is one of the most widely used MCMC algorithms. The purpose of using this algorithm is two-fold - i) to ensure that the perturbation is taking the parameter in a better region than the original location, and ii) to tune the step size so that the steps taken during perturbation are optimal. In this algorithm, the parameter perturbation $\theta^{**}$ is considered as a proposal. For this proposal, the algorithm computes two transition probabilities - one that of transitioning from $\theta^*$ to $\theta^{**}$, and the other way around from $\theta^{**}$ to $\theta^*$. Using these transition probabilities, the algorithm computes a 'metropolis acceptance ratio' $A_M(\theta^{**}, \theta^*)$ as follows:
\begin{equation}
    A_M(\theta^{**}, \theta^*) = min\left(1,\frac{p(\theta^{**}) g(\theta^*|\theta^{**})}{p(\theta^{*}) g(\theta^{**}|\theta^{*})}\right)
\end{equation}
The proposal $\theta^{**}$ is only accepted if $A_M(\theta^{**}, \theta^*) > U(0,1)$ where $U(0,1)$ is a random draw from a uniform distribution between 0 and 1. 
Of the $\theta^{**}$ that is accepted for simulation, we compute another empirical acceptance ratio - by finding out how many of these perturbed parameters achieve the distance metric $dist_{\theta^{**}}$ below the tolerance $\epsilon^i$:
\begin{equation}
    A(\theta^{**}) = \frac{num_{accepted}}{num_{attempts}}
\end{equation}
In MH, a step-size adaptation is done to maintain this empirical acceptance ratio as close to possible to the target acceptance ratio $A_T = 0.234$:
\begin{equation}
    s_{new} = s_{old} * \frac{(A(\theta^{**}) - A_T)}{(1 - A_T)(num_{accepted} + 1)}
\end{equation}
This adaptation is performed usually in the first 10\% of the ABC-SMC process, after which the rest of the process continues with a fixed step-size. This is done to avoid the step size becoming infinitesimal, which leads to the ABC-SMC process coming to a complete stop.

\section{Design}
In this section, we shall first describe how we develop a massively parallel version of the current ABC-SMC MCMC algorithm for the epidemiology model. Then, we shall discuss some of the issues encountered in this approach and the sources of incompatibility of MCMC-based step-size tuning and massively parallel simulations. Then, we shall go in-depth on the proposed algorithm which takes a novel approach on how the concept of step size can be reinvigorated for the massively parallel simulation regime. 
\subsection{Parallel ABC-SMC with MCMC Tuned Step-Size}
In earlier work \cite{kulkarni2020hardwareaccelerated,OG_ABC_IPU}, we explored the possibility of massively parallel simulations of the epidemiology model for COVID-19. We achieved up to 500k parallel 50-day simulations for the model. Using those massively parallel simulations, we show that ABC for parameter learning can be accelerated using GPUs and other emerging hardware accelerators. 

In this work, as a first step and to set up a baseline, we do the same for the current state-of-the-art ABC-SMC MCMC algorithm, i.e. to develop P-ABC-SMC MCMC. The parallelized simulation kernel used in the previous work's ABC process is used as-is.  This demonstrates the re-usability and modularity of the proposed approach. We keep the degree of parallelism to 100k, but the simulation is done for 120 days. In this work, the challenge is to perform parallelized draws of $\theta^*$ (equation 11), parallelized perturbations for $\theta^{**}$, and most importantly, parallelized MH algorithm for step-size tuning. 

The parallelized draws from the current stage's prior $\pi^i(\theta)$ were enabled by the tensorized categorical distribution. For parallelized perturbations through Gaussian random walk,  we employ the reparamterization trick similar to one performed in variational auto-encoders \cite{vae}, where all parameters are sampled from $U(0,1)$ and then transformed into their corresponding domains. 

For step-size tuning, the adaptation is done once per batch of 100k parallel simulations. During the simulation, we compute the empirical acceptance ratio which is then used to tune the step size for the next batch of 100k parallel simulations. This is done for the first 10\% runs of the P-ABC-SMC process. 

This approach provides tremendous performance gains - the MATLAB code of the non-parallelized state-of-the-art (ABC-SMC MCMC) implemented in \cite{Warne2020_github} on a Xeon CPU, reportedly takes $\sim 2$ hours, while the parallelized version developed by us (P-ABC-SMC MCMC), run on the Nvidia Tesla T4, takes only $\sim 400$ seconds, which represents a $\sim 18x$ performance gain. This provides a great baseline for the proposed algorithm, which performs even better as evidenced in later sections.

\subsection{Parallel ABC-SMC with Beta Distributed Step-Sizes}
As discussed earlier, the main contribution of the proposed algorithm is generalizing the concept of step-size in a massively parallel simulation regime. This is done by replacing the 'regular' Gaussian random walk with a MCMC tuned step size (described in the earlier section, see Equations 12,13,14,15) with a hierarchical Gaussian random walk, where step size of the random walk process itself is sampled from a tuned Beta distribution. Hence, each of the 100k simulations being run in parallel have their own unique step size, which is sampled from a Beta distribution. The resultant distribution of perturbations now contains a homogeneous mix of step sizes which efficiently explore the parameter space. In this section, we shall first go through the high-level motivation in the development of this algorithm, followed by a detailed description of the algorithm.

\subsubsection{Exploration and Exploitation in the Massively Parallel Regime}
One of the key concepts in machine learning and stochastic optimization is that of the exploration-exploitation trade-off. The goal of a ML algorithm is to find the best set of parameters for a model given data. This involves iteratively taking steps in the parameter state space. The size of the steps being taken at any given iteration of the algorithm is one of the most important aspects of the algorithm. Initially, the step-size is large to quickly find areas of lower metric (or 'loss') values. This is known as the exploration phase of the algorithm. In the later stages of the algorithm, the step size is small, to avoid moving out of the current minima and to obtain the best possible local value, known as exploitation. Hence, in the initial stages, the algorithm trades off exploitation for exploration, while in the later stages, it trades off exploration for exploitation. 

In the massively parallel regime, this exploration-exploitation trade-off is no longer required - there is enough parallelism for both. By careful design of how step sizes are allocated to the 100k simulations, a careful balance of exploration \i{and} exploitation can be achieved to get the best of both. While some of the 100k simulations explore new space, others exploit the small gains that can be achieved with small step sizes. This balance provides unique benefits throughout all stages of the ABC-SMC algorithm. The important question, then, is how to strike this balance of allocation of parallel simulations to explore or exploit (and everything in between)? This the the main intuition behind this work - this balance is achieved with the help of a tuned Beta distribution. 

\subsubsection{Beta-Distributed Step-Sizes}
The Beta distribution is a continuous probability distribution defined in the interval $[0,1]$. The shape of Beta distribution's Probability Density Function (PDF) is defined by two shape parameters denoted (confusingly so) as $Beta(\alpha, \beta)$. It is through changing these shape parameters that we can modify the Beta distribution's PDF. For the purposes of this algorithm, \textit{this modified PDF is used a allocation device which determines step sizes for each of the 100k simulations being run in parallel}. 

\begin{figure}[tbp!]
\centering
\includegraphics[width=275pt]{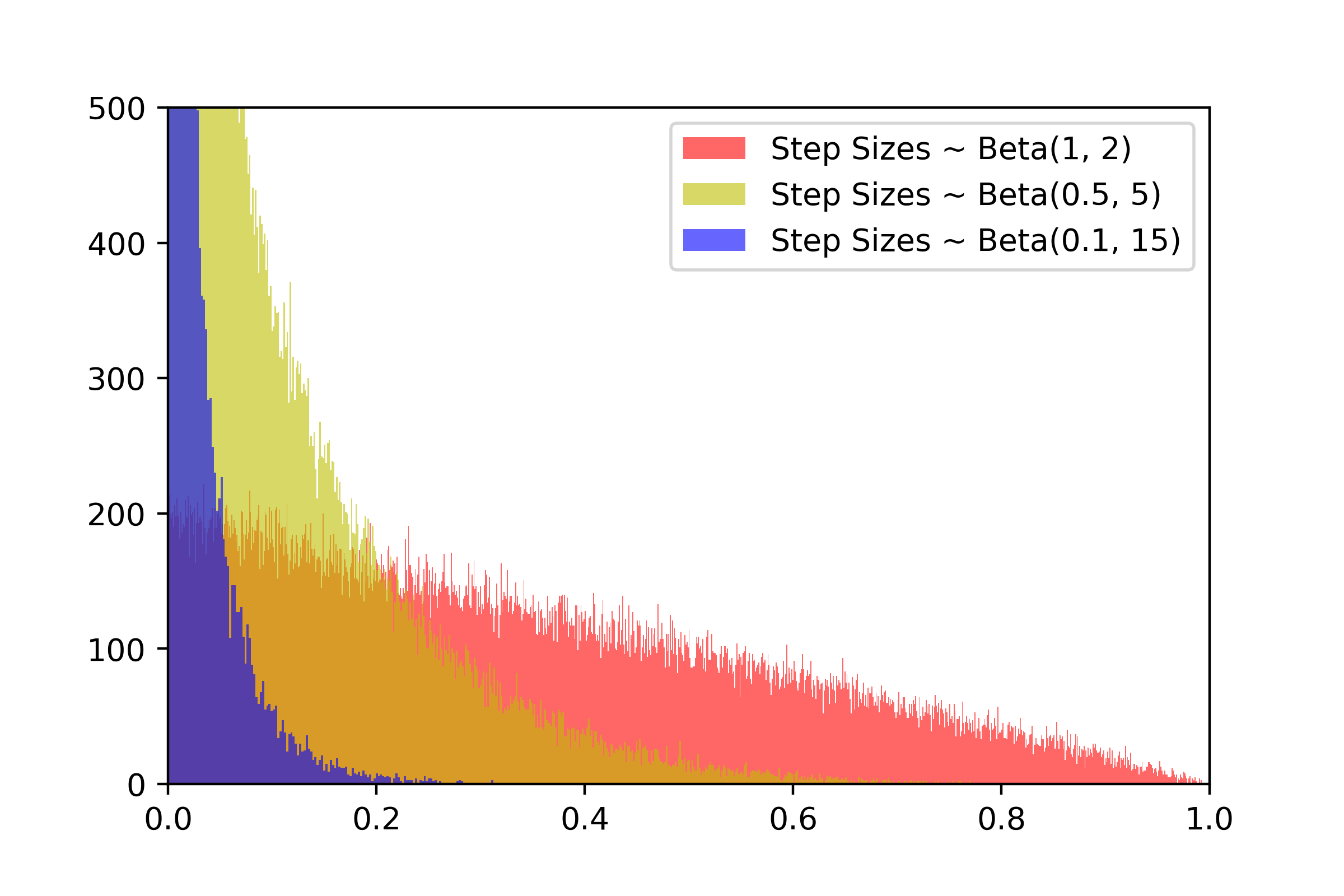}
\caption{Step-sizes sampled from Beta distribution of different shape parameters, which are representative of various stages of the P-ABC-SMC process. See Fig. \ref{f:perturbations} on how these step-size samples affect the Gaussian random walk perturbations in the various stages of the P-ABC-SMC process.}
\label{f:beta-dsitributed-step-sizes}
\end{figure}

Hence, the main reason for choosing Beta distribution instead of, for example, a truncated Gaussian, is because of how malleable the PDF is with the use of its shape parameters. These shape parameters provide an efficient way to perform the step-size allocation in a dynamic fashion. As the mass of the PDF shifts closer to the near-zero region (see Figure \ref{f:beta-dsitributed-step-sizes}), it naturally adapts to this limit. On the other hand, a truncated Gaussian does not provide the same control and its PDF does not change naturally in the near-zero region, but rather is cut-off abruptly.

\begin{figure}
     \centering
     \begin{subfigure}[b]{0.475\textwidth}
         \centering
         \includegraphics[width=\textwidth]{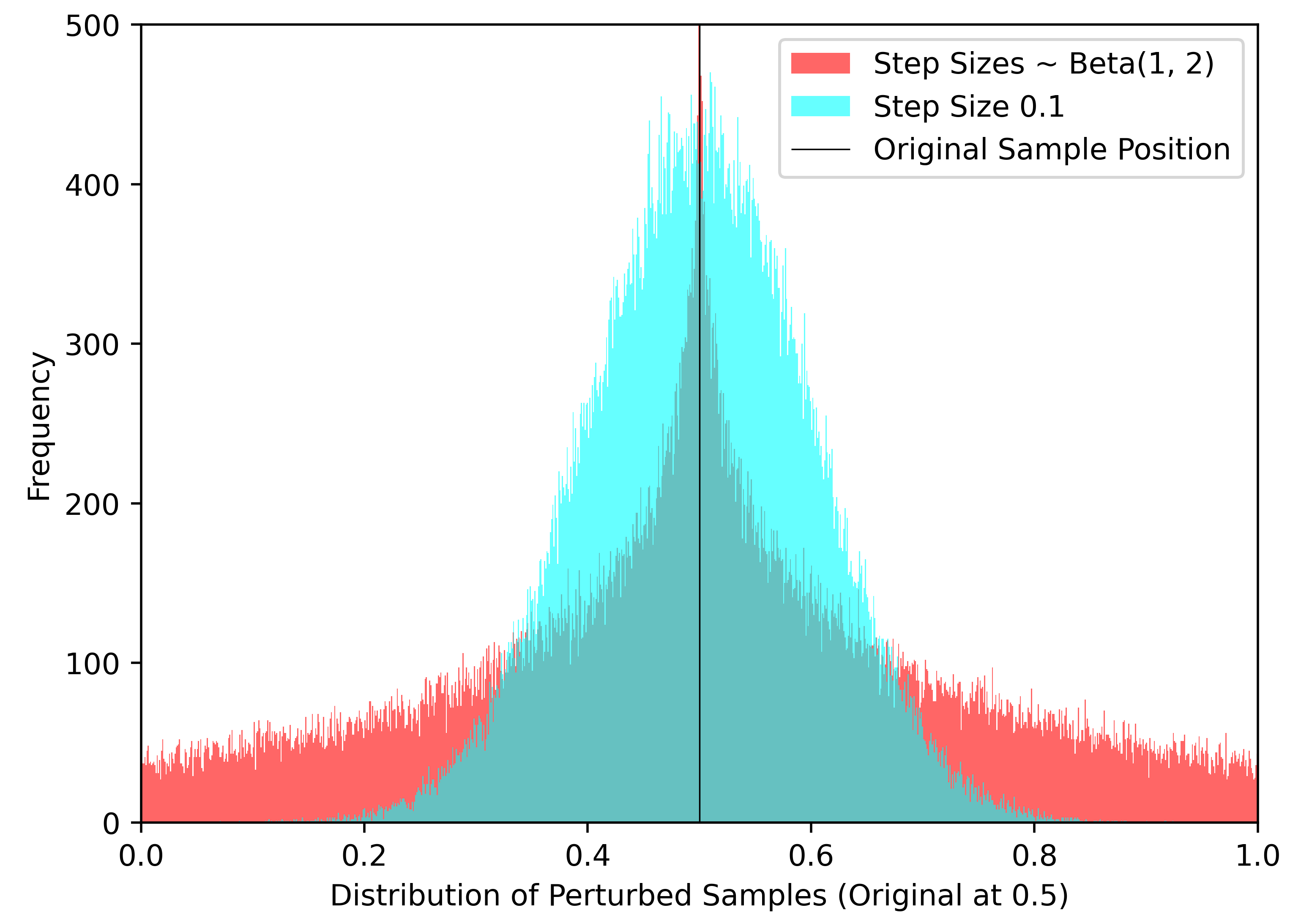}
         \caption{Initial Stage}
         \label{f:Initial Stage}
     \end{subfigure}
     \hfill
     \begin{subfigure}[b]{0.475\textwidth}
         \centering
         \includegraphics[width=\textwidth]{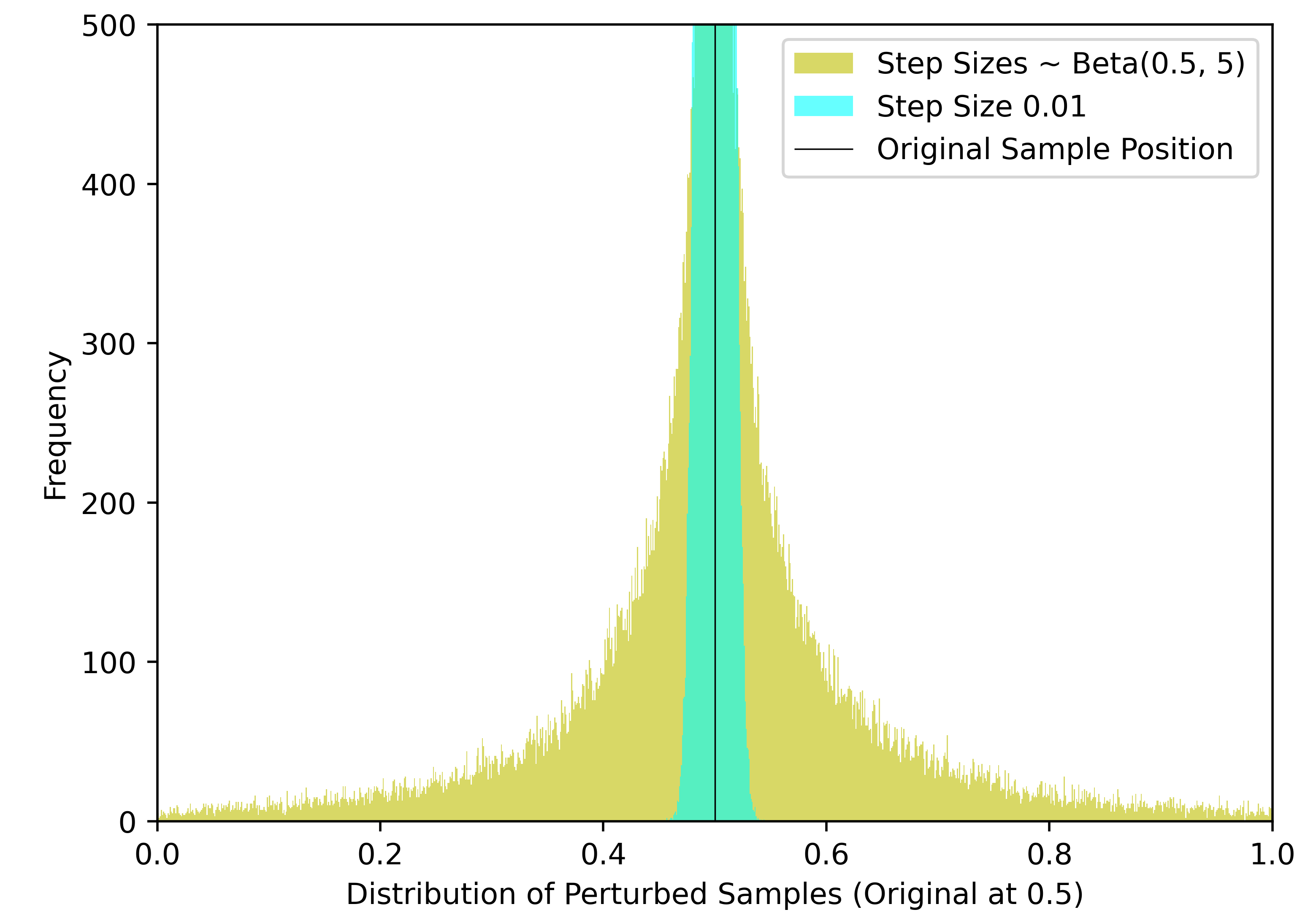}
         \caption{Intermediate Stage}
         \label{f:Intermediate Stage}
     \end{subfigure}
     \hfill
     \begin{subfigure}[b]{0.475\textwidth}
         \centering
         \includegraphics[width=\textwidth]{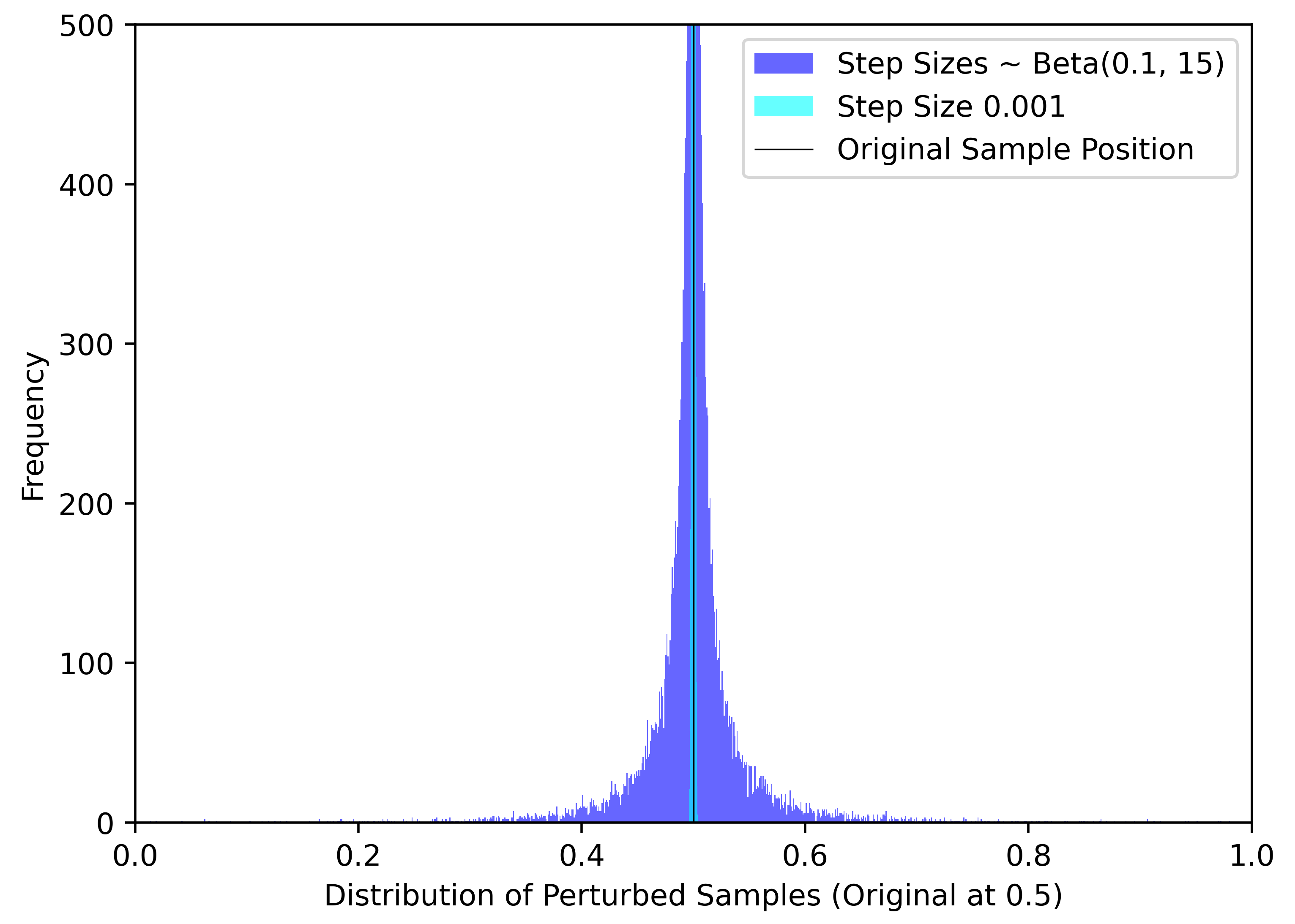}
         \caption{Late Stage}
         \label{f:Late Stage}
     \end{subfigure}
        \caption{Demonstrating the distribution of perturbed samples in three scenarios of P-ABC-SMC progression, with original position at 0.5; (a) Initial stages; (b) Intermediate stages; (c) Late stages. Plots show how while the single step-size approach clearly trades of exploration to exploitation as stages progress, the proposed Beta step-size approach provides a balance of exploration and exploitation throughout all the stages.}
        \label{f:perturbations}
\end{figure}

\begin{figure*}[htbp!]
\centering
\includegraphics[width=320pt]{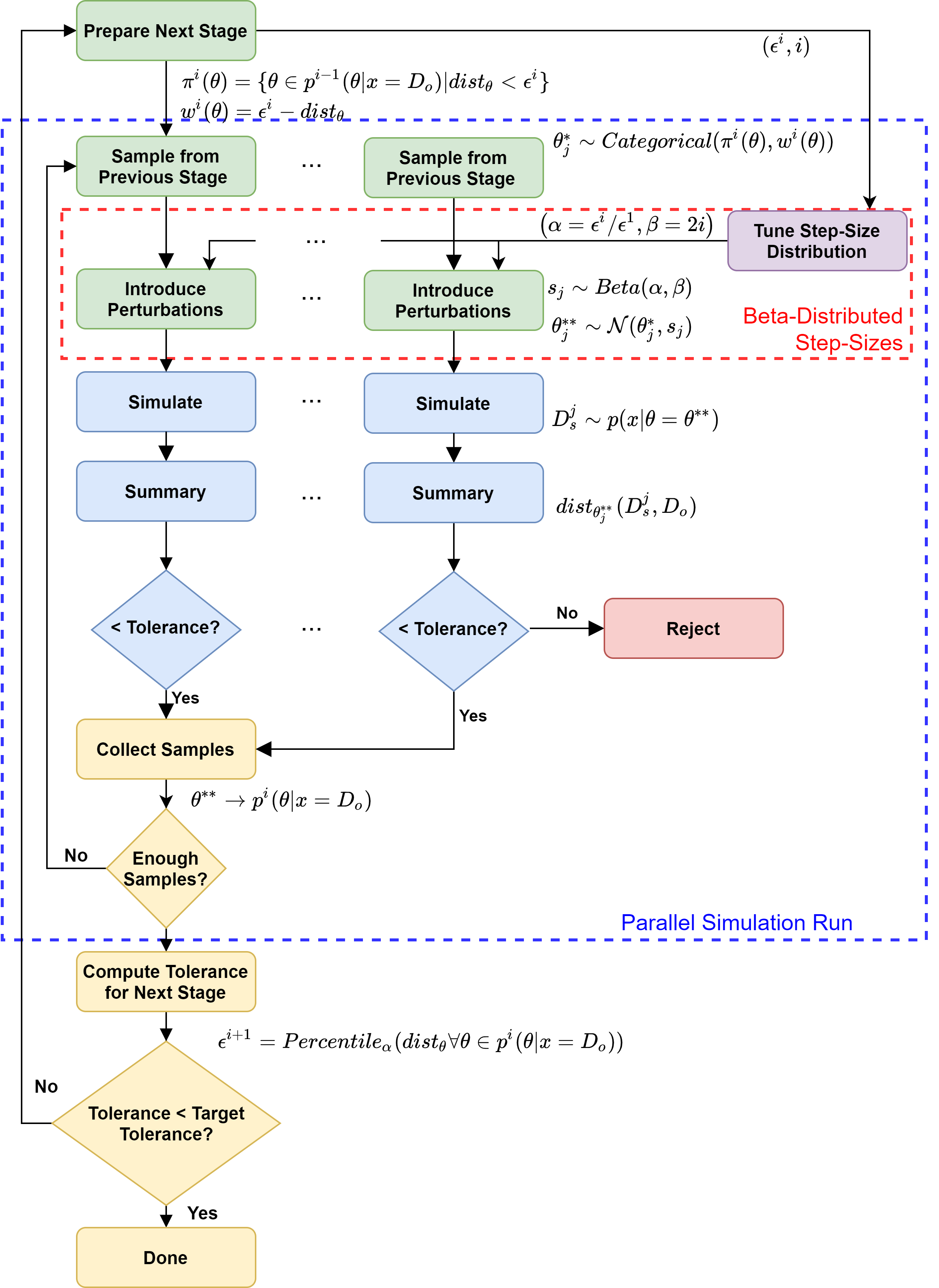}
\caption{Overview of the Parallelized ABC-SMC with Beta-Distributed Step-Sizes (P-ABC-SMC BDSS) algorithm. Highlighted in blue is the section changed from original ABC-SMC towards development of parallel ABC-SMC. Highlighted in red is the section involving the novel perturbation technique, with step-size distribution and sampling of Beta-distributed step-sizes during the perturbation step. The process of Beta distribution tuning is also shown.}
\label{f:P-ABC-SMC_BDSS_overview}
\end{figure*}

Fig. \ref{f:beta-dsitributed-step-sizes} shows 3 typical distributions of step sizes sampled from 3 different configurations of the Beta shape parameters. The shape parameters are chosen to be indicative of the initial (red), intermediate (yellow), and final (blue) stages of the ABC-SMC algorithm. In the initial stages, the idea is to allow for exploration, which requires that we allow for all possibilities of step sizes, but there should also be a slight bias toward smaller step sizes. This is captured by the $Beta(1,2)$ distribution. As the ABC-SMC process progresses, it becomes increasingly unlikely that a very large step size would yield better results, so the bias should move increasingly toward lower values and the spread should tighten. These changes are captured by the distributions $Beta(0.5,5)$ and $Beta(0.1,15)$. 

Fig. \ref{f:perturbations} shows how these step size samples lead to actual perturbations in the Gaussian random walks for a parameter sample at 0.5. To compare with the baseline algorithm, we also show how the perturbations emerge from the typical values MCMC tuned step sizes in the respective stages. It can be clearly seen that through the progression of the P-ABC-SMC algorithm, the MCMC tuned step results in perturbations that are initially moderately exploratory, but quickly become biased toward pure exploitation in the intermediate and late stages. Contrarily, the perturbations emerging from the BDSS approach, while becoming increasingly biased toward exploitation, still maintain the exploratory aspect to them thanks to the wider tail of the distributions throughout the P-ABC-SMC algorithm. 

As shall be seen in the next section, this difference in the distribution of the perturbations between the MCMC and BDSS approaches provides great benefits in terms of avoiding local minima and efficiency in the number simulations required. Although the plots in Fig. \ref{f:beta-dsitributed-step-sizes} show how the typical shapes of the Beta distribution \textit{should} look like, we still need a way to ensure the shape changes according to the P-ABC-SMC process. To ensure the effective adaption of the shape of the Beta's PDF, we employ tuning steps, which are discussed in detail now.

\subsubsection{Tuning the Beta Step-Size Distribution}
In P-ABC-SMC, the progress of the ABC-SMC algorithm is captured by two metrics - the number of stages completed, and the tolerance of the current stage. In the case of P-ABC-SMC with MCMC tuned step sizes, we did not require to take all of the above factors into consideration - the tuning was performed using the Metropolis acceptance ratio and the empirical acceptance ratio (see background section). In the P-SBC-SMC BDSS, we are not using MCMC, and we also do not perform any intermediate accept/reject steps based on the transition probabilities. 

Instead, we focus on the two metrics to track the progress of the P-ABC-SMC algorithm and use them to tune the Beta distribution. The perturbation step of the proposed algorithm is as follows:
\begin{equation}
    s^i_{j} \sim Beta\left( \alpha^i = \frac{\epsilon^i}{\epsilon^1}, \beta^i = 2i \right) \forall j \in \{1,2,\dots,100k\}
\end{equation} 
and 
\begin{equation}
    \theta^{**}_j \sim \mathcal{N}(\theta^*_j,s^i_j) \forall j \in \{1,2,\dots,100k\}
\end{equation}

Here, we can see that the shape parameters of the Beta are tuned using the the current tolerance value $\epsilon^i$ and the current stage of P-ABC-SMC process $i$. In the regime of operation $\alpha \in [0,1], \beta \in [1, \infty)$, they loosely correspond to the mean and precision (i.e. inverse variance) of the Beta distribution. Hence, by setting $\alpha^i = \frac{\epsilon^i}{\epsilon^1}$, we move the mean of the step size distribution closer to 0 as we keep getting better tolerances. At the same time, as the number of stages increases, $\beta^i = 2i$ also increases, effectively decreasing the variance. The new P-ABC-SMC BDSS algorithm is summarized in Fig. \ref{f:P-ABC-SMC_BDSS_overview}

There are two other methods of parametrization in a Beta distribution - the mean-variance and the mode-concentration parameterizations. We try both of these, but the original $\alpha, \beta$ parameterization performs the best. 

\subsection{Theoretical Correctness and Convergence Guarantees for ABC-SMC}

 Likelihood-based MCMC methods such as Gibbs' Sampling or Metropolis-Hastings\cite{MHAlgo}, give asymptotic guarantees of converging to the true posterior of the model as the number of samples obtained approaches infinity. This guarantee is in part based on the correctness of the likelihood function.
 
 On the other hand, like all SBI methods, ABC-SMC is likelihood-free. In these methods, the likelihood function is approximated by a distance metric, making these algorithms inherently approximate. To accept samples in the approximate posterior, these algorithms have an artificial cut-off of the tolerance value $\epsilon$. Hence, the only case of correctness that can be made for SBI algorithms is that as this tolerance value approaches zero, the approximate posterior converges to the true posterior. There are no asymptotic guarantees though; even after an infinite number of ABC-SMC steps, the tolerance is not guaranteed to approach zero. Hence, ABC-SMC algorithms are more accurate when they can manage to accept approximate posterior samples at lower tolerance values. This is applicable to all ABC-SMC algorithms, including the two we test in this work.
 
 Both MCMC-based and BDSS approaches to ABC-SMC differ in methods of proposing samples to the ABC step; the actual acceptance and determination of the tolerance for the next stage is done in the same way. Hence we believe that for both of these algorithms the same guarantees hold.

\begin{figure*}[htbp!]
     \centering
     \begin{subfigure}[b]{0.45\textwidth}
         \centering
         \includegraphics[width=\textwidth]{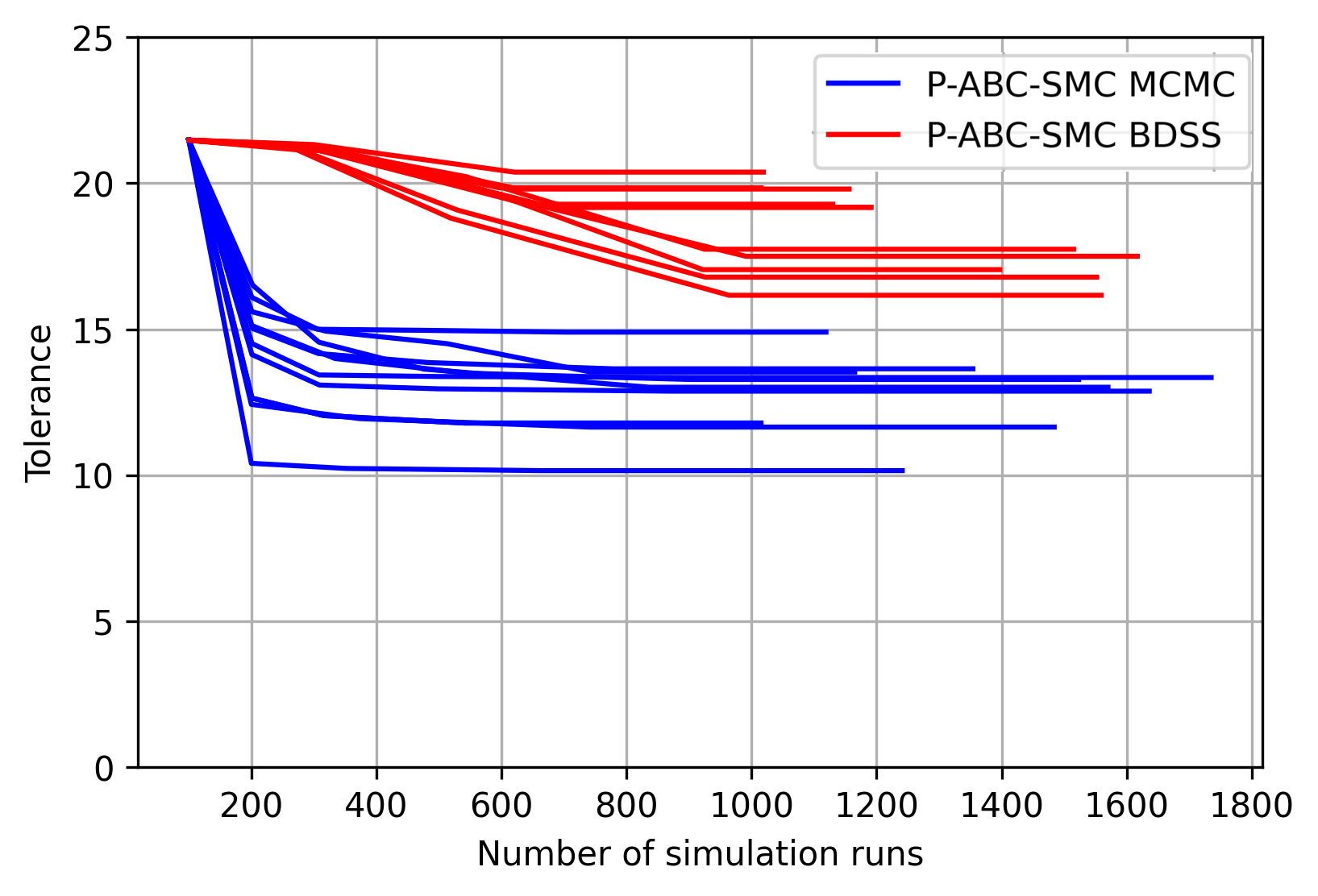}
         \caption{Simulations per run: 10}
         \label{f:10_parallel}
     \end{subfigure}
     \begin{subfigure}[b]{0.45\textwidth}
         \centering
         \includegraphics[width=\textwidth]{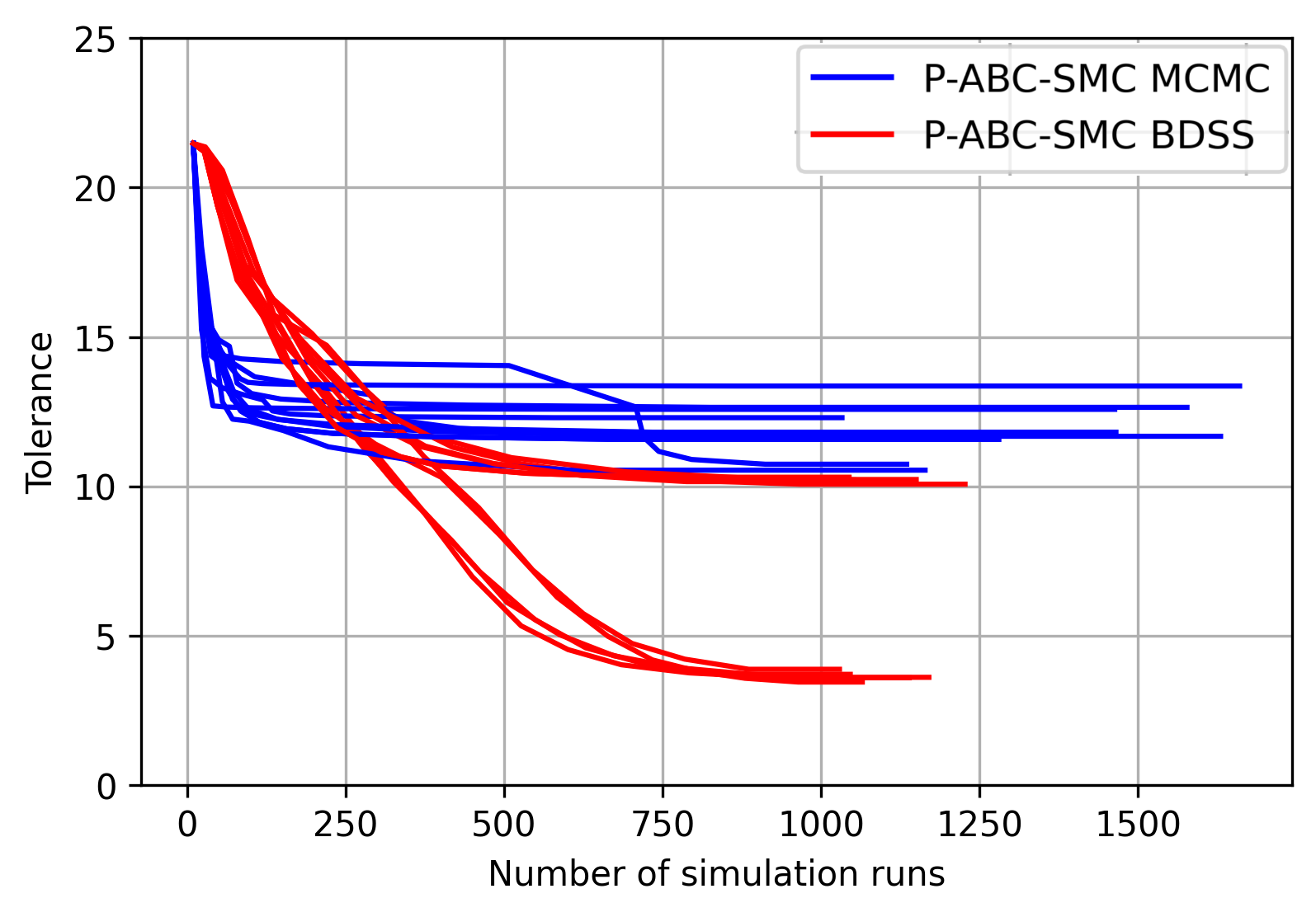}
         \caption{Simulations per run: 100}
         \label{f:100_parallel}
     \end{subfigure}
     \begin{subfigure}[b]{0.45\textwidth}
         \centering
         \includegraphics[width=\textwidth]{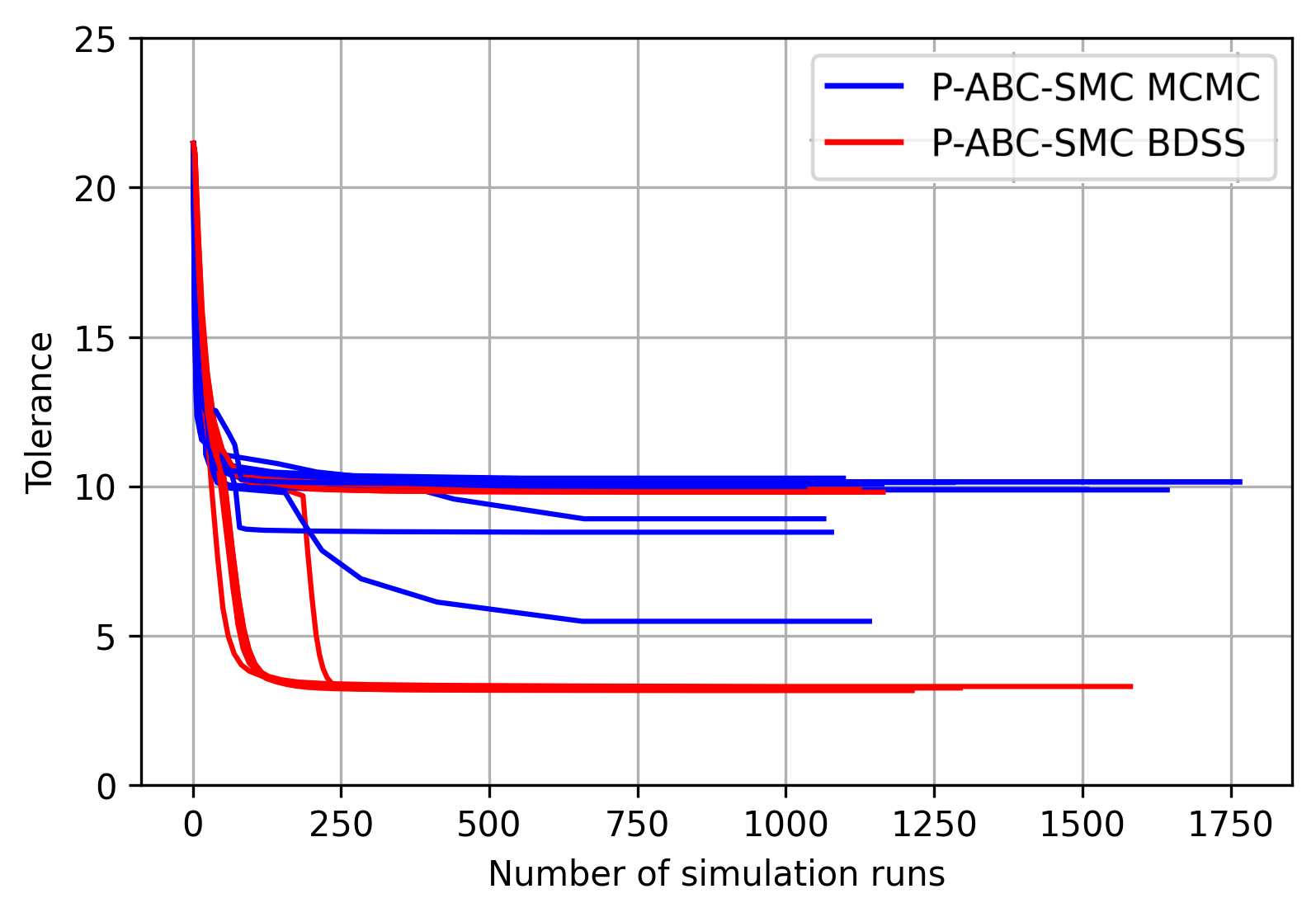}
         \caption{Simulations per run: 1000}
         \label{f:1000_parallel}
     \end{subfigure}
     \begin{subfigure}[b]{0.45\textwidth}
         \centering
         \includegraphics[width=\textwidth]{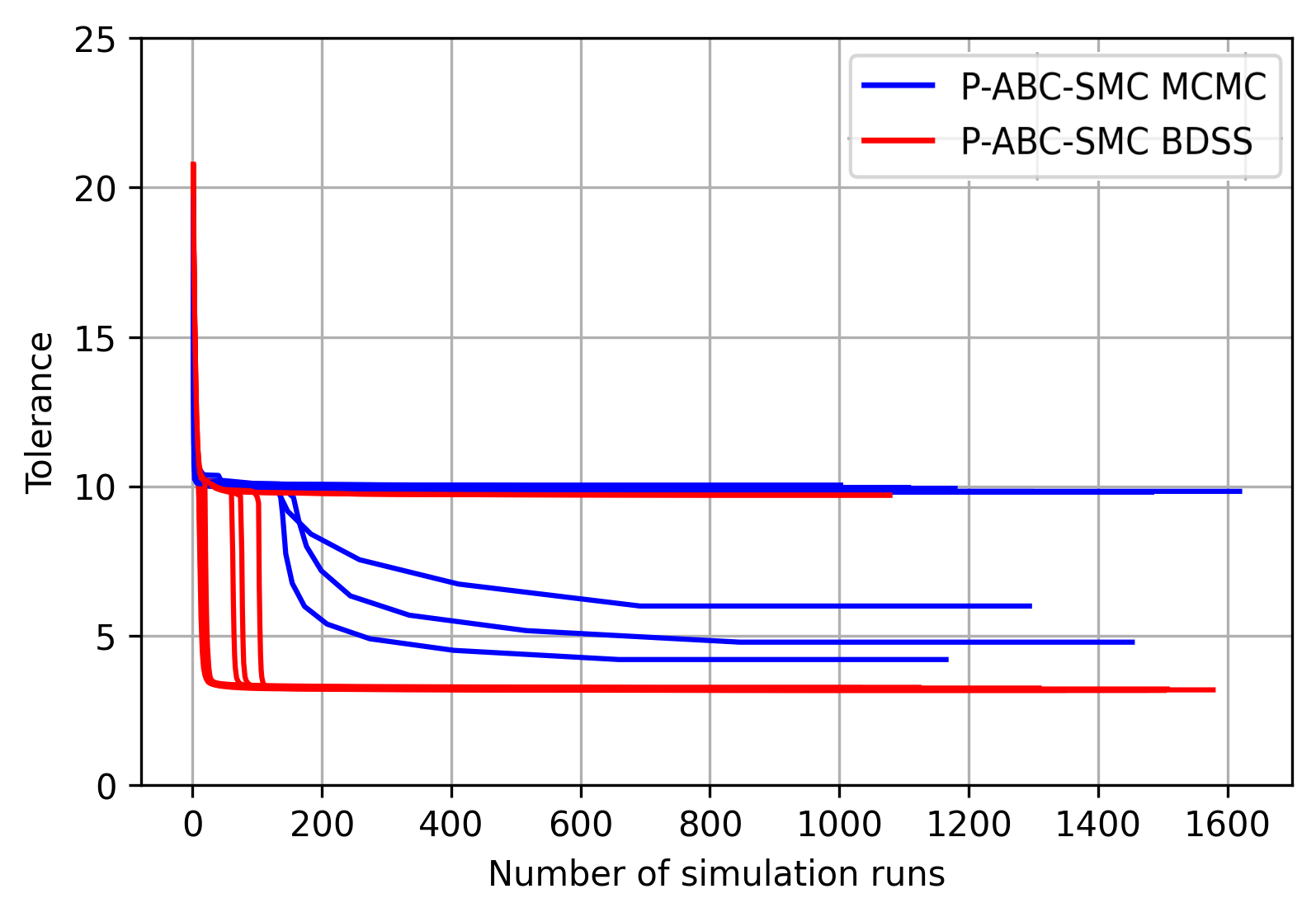}
         \caption{Simulations per run: 10000}
         \label{f:10000_parallel}
     \end{subfigure}
     \begin{subfigure}[b]{0.55\textwidth}
         \centering
         \includegraphics[width=\textwidth]{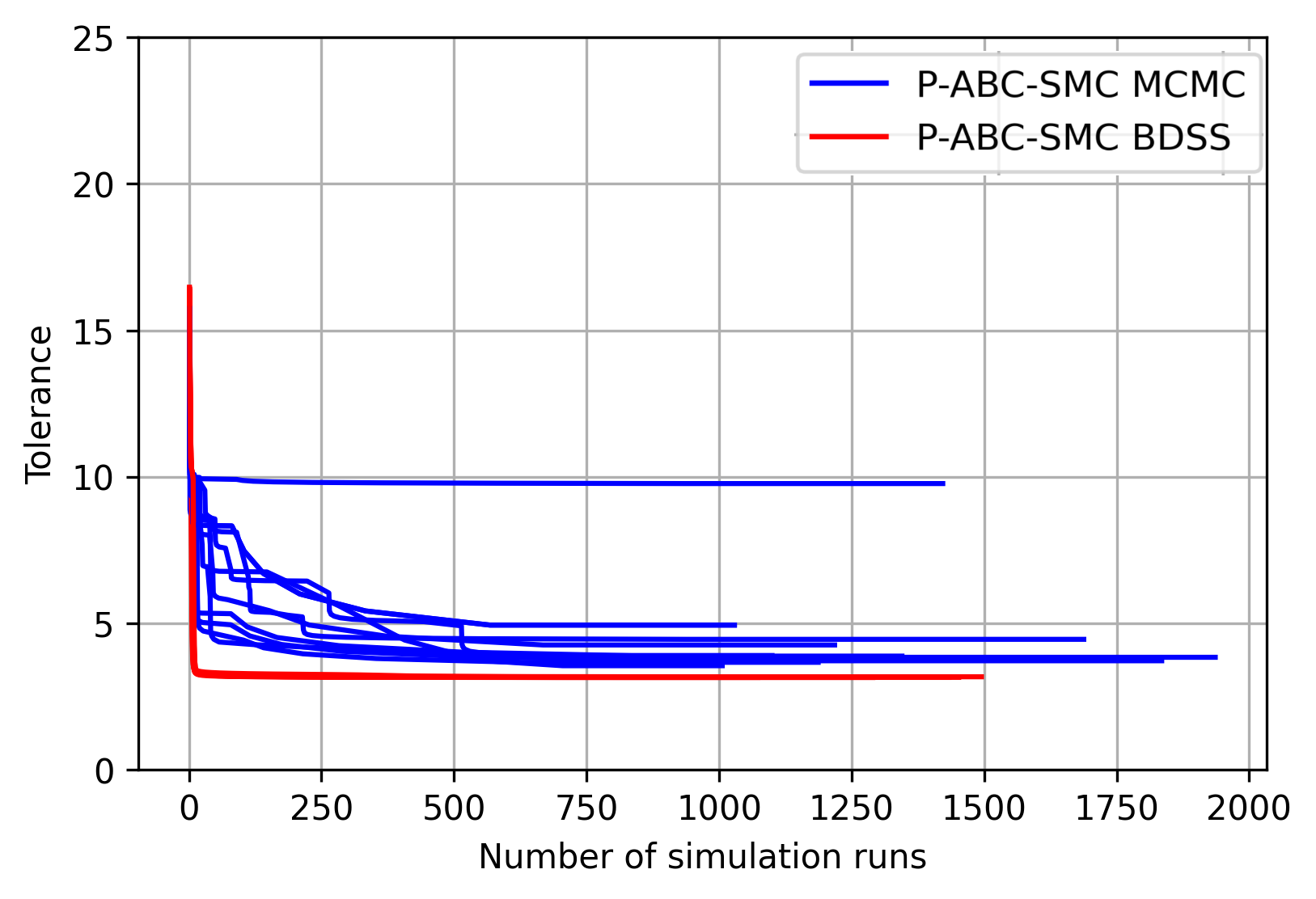}
         \caption{Simulations per run: 100000}
         \label{f:100000_parallel}
     \end{subfigure}
        \caption{Effectiveness of P-ABC-SMC BDSS with increasing degree of parallelism, compared to P-ABC-SMC MCMC. Experiment run for 120-day case data of Italy, over 10 independent trials limited to the stage which reached 1000 simulation runs. (a) When only 10 parallel simulations are performed per run, the MCMC based approach performs much better than BDSS; (b) At 100 simulations per run, the benefits of BDSS are visible - in 4 of the 10 trials, the algorithm manages to break out of the local mimima; (c) At 1000 simulations per run, we see BDSS has much faster convergence to lower minima in most of the trials, while the three trials of the MCMC approach that manage to break off from local minima still have slow convergence; (d) At 10000 simulations per run, all but one trials for BDSS quickly converge to the minima, while the in the MCMC approach the convergence is much slower; (e) for 100k simulations per run, the BDSS approach converges consistently to the minima within 10 simulation runs, while the MCMC approach still has one trial stuck in local minima, and the rest converging 100 times slower and to higher minimas than the proposed.}
        \label{f:parallelsim_comparison}
\end{figure*}

Given the lack of theoretical guarantees, we instead validate these algorithms on the ability of approximate posterior samples accepted by them to predict unseen data. This is described in more detail in Section 4.2.
\section{Experiments and Results}
To test the baseline and proposed algorithms, we use them to learn parameters for the stochastic epidemiology model (section 2.1). The model has a set of 8 parameters that governs how an epidemic spreads through a population of a nation. The goal of the parameter learning process is to obtain the parameters with the least value of distance metric from the real-world case data. For the purposes of this work, we trained on 120-day case data for Italy. 

The algorithms are evaluated on three metrics - best tolerance achieved, simulation efficiency, consistency in run-to-run variance. The best tolerance achieved demonstrates the quality of samples achieved - lower the tolerance, better the parameters fit to real-world data. The simulation efficiency is measured by how many simulation runs are required before the algorithm reaches it's best possible tolerance. The run-to-run variance describes how consistent the algorithm is, both in lowest tolerance achieved as well as the simulation efficiency. 

Finally, we shall also evaluate the the learnt parameters in their ability to predict future cases. This is done by using the 120-day data to learn the parameters, and then using those learnt parameters to recreate the case data for those 120 day window as well as 30 additional days. These 30 days act like the test data set and allow us to measure the accuracy of the parameters. 

\subsection{Performance Comparison}

\subsubsection{Experiment Setup}
To compare the new P-ABC-SMC BDSS with the baseline P-ABC-SMC MCMC, we perform the task of learning model parameters from 120-day case data of Italy. We conduct 10 independent trials to gauge the run-to-run variance of both the algorithms. These trials are conducted for 5 different levels of parallelism - 10, 100, 1k, 10k, and 100k simulations per run, respectively. The number of samples to be collected is set to 1000. For consistency, we establish a stopping criterion - the P-ABC-SMC inference process stopped on the stage which reached 1000 simulation runs. Since there could still be additional simulation runs required for that stage to finish, the actual number of runs varies from 1000 to 1900. At each stage, we record the tolerance achieved by each of the algorithms. This allows us to compare the quality of the samples (tolerance achieved) and the number of simulation runs required to obatain them. Hence, with these two metrics, we generate 5 plots, one for each level of parallelism in Fig \ref{f:parallelsim_comparison}.

\begin{figure*}[!ht]
     \centering
     \begin{subfigure}[b]{0.45\textwidth}
         \centering
         \includegraphics[width=\textwidth]{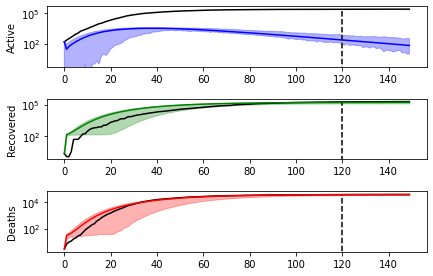}
         \caption{Predictive Simulation from P-ABC-SMC MCMC in 10 simulation runs}
         \label{f:10_runs_mcmc}
     \end{subfigure}
     \hfill
     \begin{subfigure}[b]{0.45\textwidth}
         \centering
         \includegraphics[width=\textwidth]{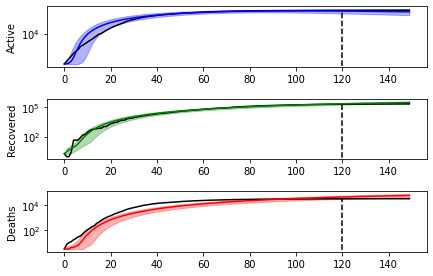}
         \caption{Predictive Simulation from P-ABC-SMC BDSS in 10 simulation runs}
         \label{f:10_runs_proposed}
     \end{subfigure}
     \hfill
     \begin{subfigure}[b]{0.45\textwidth}
         \centering
         \includegraphics[width=\textwidth]{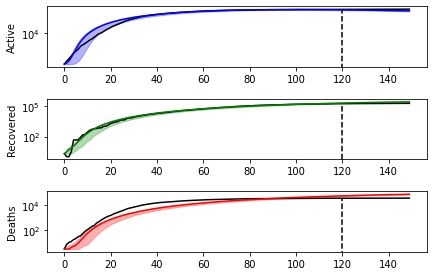}
         \caption{Predictive Simulation from P-ABC-SMC MCMC in 1000 simulation runs}
         \label{f:1000_runs_mcmc}
     \end{subfigure}
     \hfill
     \begin{subfigure}[b]{0.45\textwidth}
         \centering
         \includegraphics[width=\textwidth]{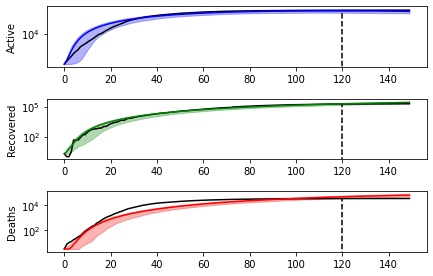}
         \caption{Predictive Simulation from P-ABC-SMC BDSS in 1000 simulation runs}
         \label{f:1000_runs_proposed}
     \end{subfigure}
        \caption{Comparing the predictive simulations of case-data using learned parameter distributions in early stages(a,b) and late stages (c,d) for P-ABC-SMC MCMC and P-ABC-SMC BDSS. Real world case data plotted in black. Training performed on 120-day case data. Predictive simulations of case-data are re-created from day zero. Plots are median (solid color lines) with 99th percentile bounds (light color bands). }
        \label{f:prediction_comparison}
\end{figure*}

\subsubsection{Results}

Fig. \ref{f:parallelsim_comparison} provides a detailed visualization of the comparison between the two approaches. When the degree of parallelism is low (see Fig. \ref{f:10_parallel}) with only 10 simulations per run, the P-ABC-SMC MCMC approach is clearly superior, as the proper tuning of the individual step size yields better results than obtaining 10 samples from a Beta distribution with wide tails.

With increasing degree of parallelism, though, the P-ABC-SMC BDSS starts to show its benefit. At just 100 simulations per run (see Fig \ref{f:100_parallel}), we can observe that while P-ABC-SMC MCMC observes substantial benefits, P-ABC-SMC BDSS performs even better. Half of the trials manage to perform comparable to P-ABC-SMC MCMC, the other half comfortably outperform, eventually approaching the best tolerance values achieved by even the later experiments with higher degree of parallelism, though they take the entirety of their allocated simulation-runs budget to do so.

With the number of simulations per run up to 1000 (see Fig \ref{f:1000_parallel}), we observe that 3 of the 10 P-ABC-SMC MCMC trials break the apparent local minima at the tolerance value of $\sim 10$, and begin to converge toward lower values, though never converging fully. On the other hand, P-ABC-SMC BDSS manages to converge to the best achieved tolerance in 6 of the 10 runs, and does so in just 250 simulation runs instead of the 1000-run limit.

The performance gap widens further with 10,000 simulations per run (see Fig \ref{f:10000_parallel}), Where the performance gain of P-ABC-SMC MCMC is only slight, while P-ABC-SMC BDSS manages to converge fully in 9 of the 10 trials, and all of them converge in less than 150 simulation runs. 

Finally, at 100k simulations per run (see Fig \ref{f:100000_parallel}), we see benefits in both P-ABC-SMC MCMC and BDSS, although the performance of the BDSS approach is vastly better. All 10 trials consistently reach close to the best tolerance in a span of just 10-20 simulation runs. P-ABC-SMC with MCMC slowly approaches the best achieved tolerance by the proposed algorithm, but the process stops short in all trials. One of the trial still remained stuck at the local minima. 

Hence, it is clear that the BDSS approach is clearly superior to the ABC-SMC MCMC approach, especially when employed in the massively parallel regime of P-ABC-SMC, for which it was developed. It can also be seen that while it performs best in degrees of parallelism of 100k, it is also beneficial to use in the regime of parallelism as low as 100 as that is when it starts to show benefits over P-ABC-SMC MCMC. The reasons for these results and their implications are discussed further in the  next section. 

P-ABC-SMC BDSS achieves a lower average tolerance value of $3.44$ in $10$ runs than the P-ABC-SMC MCMC can achieve in $\sim 1300$ runs, which is $4.01$. Hence, we report that the proposed algorithm utilizes $\sim 100 \times$ fewer simulations to achieve better tolerance values. As shown in Section \ref{sec:runtime} Moreover, at just $10$ runs, the variance of the tolerance across the $10$ independent trials is just $0.002$, while the variance of P-ABC-SMC MCMC at $\sim 1300$ runs is $0.16$.\footnote{ These values are computed by excluding the outlier with $9.7$ tolerance in case of P-ABC-SMC MCMC in Fig. \ref{f:100000_parallel}.}.

\subsection{Comparing Predictive Simulations from Learned Parameters}


\subsubsection{Experiment Setup}

We use the parameters learnt from the 120-day training data are used to simulate the case-data of those 120 days and an additional 30 days. The case data in question is the number of confirmed active (A), confirmed recovered (R), and confirmed deaths (D) for Italy. If the learnt parameters are good, the simulated case data should closely resemble the real-world case data. It is important to note that the similarity of simulated case data should be considered as being 'samples from the same underlying distribution' rather than a direct match to the real-world data. The purpose of learning the parameter distributions of the epidemiology model is to capture the underlying natural process of the epidemic, so that the simulated data from all posterior samples is the model's way of capturing all realistic trajectories the case-data could take, given the parameters we learnt from the one real-world trajectory of case data we observed. 

One of the key strengths of the proposed algorithm is its efficiency in the number of simulations required to obtain good solutions in the parameter space. This is exemplified in the experiments performed here. We create simulated the 120-day case-data for Italy using 'snapshots' of intermediate posterior parameters when both algorithms are 10 and 1000 runs into the ABC-SMC process respectively. The degree of parallelism is set to 100k simulations per run. The number of samples accepted is 1000. For each algorithm, we plot the median along with the $99^{th}$ percentile range of the case-data simulations generated from those 1000 parameter samples. For reference, we also plot the real-world case data in black.

\begin{figure}[htp!]
\centerline{\includegraphics[width=\linewidth]{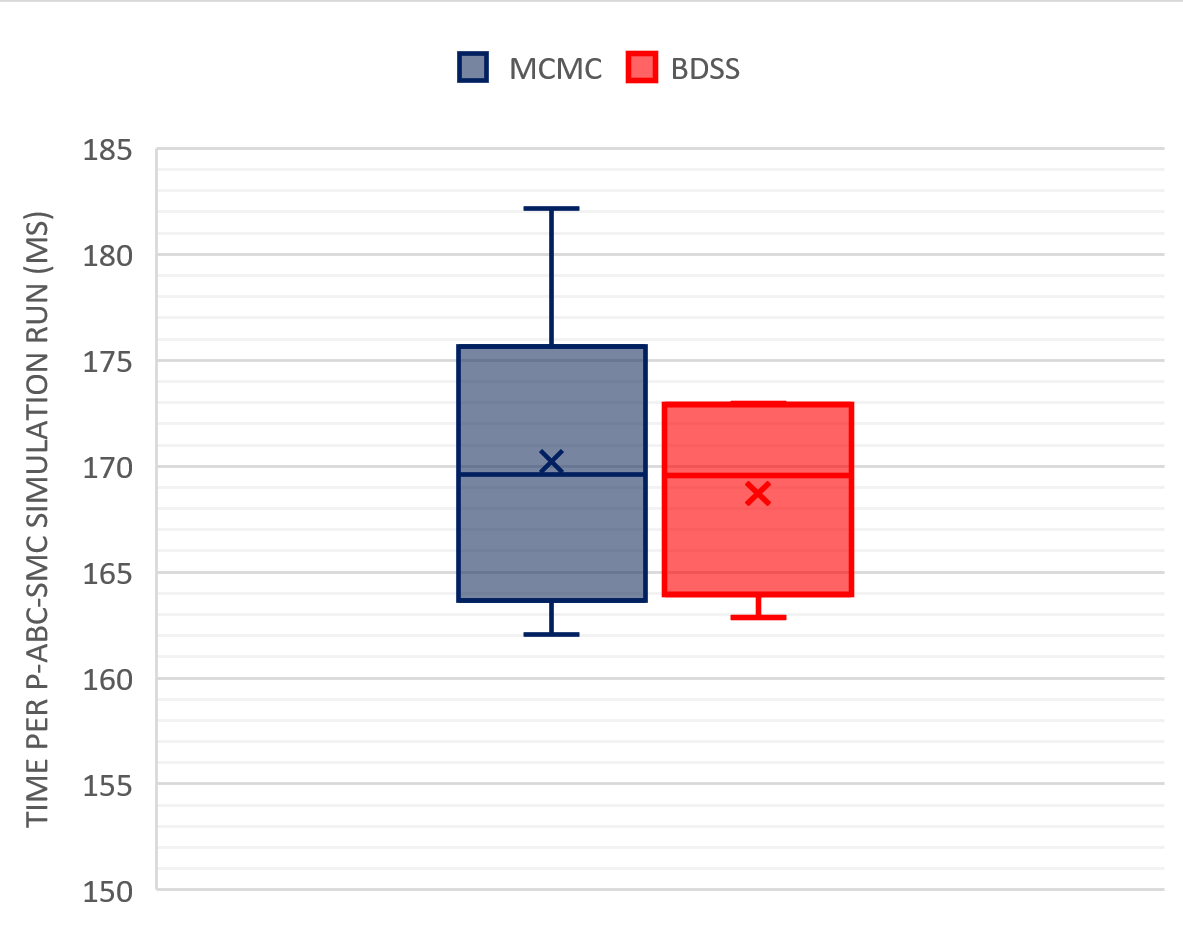}}
\caption{Runtime Comparison between P-ABC-SMC MCMC and BDSS. The plot shows the distribution of per-simulation runtimes of both algorithms, for a batch size of 100k, across 10 trials. The runtimes are very close to each other, with median values around $\sim 170$ms. The MCMC runtime has a slightly higher variance. The X's denote the mean runtimes.
}
\label{f:runtime-analysis}
\end{figure}

\subsubsection{Results}

Fig. \ref{f:prediction_comparison} shows plots generated from these experiments. As seen in Fig. \ref{f:10_runs_mcmc} the parameters obtained by P-ABC-SMC MCMC algorithm in 10 simulation runs are not even close to converge and do not generate simulated case data close to the real-world case-data for Italy. On the other hand, already at 10 simulation runs, P-ABC-SMC BDSS generates simulated case-data that shows a really good fit to real-world case-data (See Fig. \ref{f:10_runs_proposed}). 

Fig. \ref{f:1000_runs_mcmc} and \ref{f:1000_runs_proposed} show how the simulated case-data for P-ABC-SMC MCMC and P-ABC-SMC BDSS using parameters at 1000 simulation runs. Here we see that both are now producing accurate simulations of case data that closely match the real-world case data. Though we can also observe that P-ABC-SMC BDSS is generating comparable plots in just 10 simulation runs \ref{f:10_runs_proposed}. Hence, the proposed algorithm can be terminated much sooner to obtain the same (or better) quality result than P-ABC-SMC MCMC. 

\subsection{Runtime Comparison}\label{sec:runtime}
The experiments were run on a Tesla T4 GPU on the Google Colaboratory \footnote{https://colab.research.google.com/}. The goal here is to compare the relative runtime performance of the two algorithms, rather than to observe the absolute best performance numbers which could be achieved by using higher-end GPUs. We observe no significant difference in performance across the two ABC-SMC implementations. In both the P-ABC-SMC MCMC and BDSS, we observe an average of $\sim170ms$ per 100k simulation run (see Figure \ref{f:runtime-analysis}). Hence, with a similar runtime, the simulation efficiency benefits described in 4.1 directly translate to faster runtime, as P-ABC-SMC BDSS achieves better tolerance values in 10 simulation runs than P-ABC-SMC MCMC does in $\sim 1300$ runs.

We also observe that in both cases, around $50\%$ of the runtime is spent on the actual simulations (i.e., the ABC part of ABC-SMC), while the other half is spent in all the other aspects, such as resampling from previous stage, MCMC/BDSS Step tuning, perturbations, accept-reject, sample post processing etc. Specifically, P-ABC-SMC MCMC reported $84ms$ of time for actual simulations, while BDSS reported $84.5ms$.

\section{Discussion and Conclusion}

In this work, we introduce a novel algorithm that is built specifically for utilizing massively parallel hardware architectures such as GPUs to enable $\sim 100 \times$ fewer simulations than the current state of the art, while providing better quality and more consistent results across independent trails. The key contribution is the algorthimzation of step-size allocation across arbitrarily large parallel optimization processes. This algorithm provides a great avenue of utilizing the widely available GPU computing resources for scientific simulation models and cutting-edge simulation-based inference techniques such as ABC-SMC.

\subsection{Effectiveness of the Proposed Algorithm in Massively Parallel Regime}

When we consider the regime of massively parallel simulations, the P-ABC-SMC with the MCMC based approach has two main limitations. The notion of tuning a single step size along the process, and then sampling a large number of perturbations in the parameter space with a single step size is inefficient. Once the step size gets tuned to lower values with MCMC, the optimization process slows down. This leads to scenarios where the P-ABC-SMC process is stuck in a local minima and also leads to inconsistent results across independent trails. 

On the other hand, P-ABC-SMC BDSS, through the tuned Beta distribution over step sizes, introduces a much wider variety of parameter perturbations across the parallel simulations. This causes the P-ABC-SMC algorithm to obtain a wider range of possible good parameter values, which in turn leads to better results faster. We also see that through the tuning process, the proposed algorithm produces highly consistent results across independent trials, which are better than P-ABC-SMC MCMC, and in much smaller number of simulation runs. Hence, through the ABC-SMC process, the parallel simulations in hardware (P) and the proposed beta-distributed step-size allocation algorithm (BDSS) form a much better approach to accelerate simulation-based inference for scientific simulation models (P-ABC-SMC BDSS). The effectiveness of this algorithm is exemplified in Fig. \ref{f:prediction_comparison}, which shows that this algorithm has to be run only for 10 runs to obtain similar or better quality results than the 1000+ runs for P-ABC-SMC MCMC. The obtained results are also more consistent, with $\sim 80 \times$ lower variance across independent trials.

\subsection{Wider Applicability of Proposed Algorithm}
While we demonstrate the effectiveness of this algorithm by performing the novel ABC-SMC inference on a specific compartmental epidemiological model for COVID-19, we believe that the benefits may also translate to the use of this novel ABC-SMC in other scientific models, as the only major change required would be the development of the parallelized version of the scientific simulation model which we wish to learn the parameter distributions for. This can be obtained via the 'parallelization-through-tensorization' approach detailed in \cite{OG_ABC_IPU,kulkarni2020hardwareaccelerated}. The key idea is to take a single execution trace of the scientific model's simulation for a specific parameter configuration, and then to add an additional dimension (i.e., tensorize) this trace to now perform the same simulation trace for an arbitrarily large number of parameter configurations. These scientific simulation models, parallelized in this fashion, could be from a vast range of scientific domains, from elementary particle physics\cite{CMS_CERN} to cosmology\cite{abc_cosmology,CMBSim} (And everything in between \cite{sim_infer_survey}).   

Furthermore, the results obtained in this work are promising not only for the ABC-SMC algorithm, but also for stochastic optimization algorithms in general. It is theoretically possible to generalize the concept of parallel step-size allocation through a tuned step-size distribution to other stochastic optimization algorithms such as gradient descent, where the size of the step taken in direction of the gradient can be sampled from a distribution instead of being a single value.


%

\ifCLASSOPTIONcompsoc
  \section*{Acknowledgments}
\else
  \section*{Acknowledgment}
\fi

The authors would like to thank Facebook Research for their 'Probability and Programming' grant to support this work.

\ifCLASSOPTIONcaptionsoff
  \newpage
\fi



\bibliographystyle{IEEEtran}
\bibliography{references.bib}

\begin{thebibliography}{10}
\providecommand{\url}[1]{#1}
\csname url@samestyle\endcsname
\providecommand{\newblock}{\relax}
\providecommand{\bibinfo}[2]{#2}
\providecommand{\BIBentrySTDinterwordspacing}{\spaceskip=0pt\relax}
\providecommand{\BIBentryALTinterwordstretchfactor}{4}
\providecommand{\BIBentryALTinterwordspacing}{\spaceskip=\fontdimen2\font plus
\BIBentryALTinterwordstretchfactor\fontdimen3\font minus
  \fontdimen4\font\relax}
\providecommand{\BIBforeignlanguage}[2]{{%
\expandafter\ifx\csname l@#1\endcsname\relax
\typeout{** WARNING: IEEEtran.bst: No hyphenation pattern has been}%
\typeout{** loaded for the language `#1'. Using the pattern for}%
\typeout{** the default language instead.}%
\else
\language=\csname l@#1\endcsname
\fi
#2}}
\providecommand{\BIBdecl}{\relax}
\BIBdecl

\bibitem{CMS_CERN}
C.~Collaboration, S.~Chatrchyan, G.~Hmayakyan, V.~Khachatryan, A.~Sirunyan,
  W.~Adam, T.~Bauer, T.~Bergauer, H.~Bergauer, M.~Dragicevic \emph{et~al.},
  ``The cms experiment at the cern lhc,'' 2008.

\bibitem{abc_biochem}
J.~M. Tomczak and E.~Wglarz-Tomczak, ``Estimating kinetic constants in the
  michaelis--menten model from one enzymatic assay using approximate bayesian
  computation,'' \emph{FEBS letters}, vol. 593, no.~19, pp. 2742--2750, 2019.

\bibitem{Warne2020}
D.~J. Warne, A.~Ebert, C.~Drovandi, A.~Mira, and K.~Mengersen, ``{Hindsight is
  2020 vision: Characterisation of the global response to the COVID-19
  pandemic},'' \emph{medRxiv}, p. 2020.04.30.20085662, may 2020.

\bibitem{abc_cosmology}
A.~Weyant, C.~Schafer, and W.~M. Wood-Vasey, ``Likelihood-free cosmological
  inference with type ia supernovae: approximate bayesian computation for a
  complete treatment of uncertainty,'' \emph{The Astrophysical Journal}, vol.
  764, no.~2, p. 116, 2013.

\bibitem{Galactic_simulations}
\BIBentryALTinterwordspacing
R.~E. Sanderson, A.~Wetzel, S.~Loebman, S.~Sharma, P.~F. Hopkins,
  S.~Garrison-Kimmel, C.-A. Faucher-Giguère, D.~Kereš, and E.~Quataert,
  ``Synthetic gaia surveys from the fire cosmological simulations of milky
  way-mass galaxies,'' \emph{The Astrophysical Journal Supplement Series}, vol.
  246, no.~1, p.~6, Jan 2020. [Online]. Available:
  \url{http://dx.doi.org/10.3847/1538-4365/ab5b9d}
\BIBentrySTDinterwordspacing

\bibitem{abc_cognitive}
A.~Kangasr{\"a}{\"a}si{\"o}, J.~P. Jokinen, A.~Oulasvirta, A.~Howes, and
  S.~Kaski, ``Parameter inference for computational cognitive models with
  approximate bayesian computation,'' \emph{Cognitive Science}, vol.~43, no.~6,
  p. e12738, 2019.

\bibitem{abc_econometrics}
L.~E. Calvet and V.~Czellar, ``Accurate methods for approximate bayesian
  computation filtering,'' \emph{Journal of Financial Econometrics}, vol.~13,
  no.~4, pp. 798--838, 2015.

\bibitem{abc_evolution}
M.~A. Beaumont, ``Approximate bayesian computation in evolution and ecology,''
  \emph{Annual review of ecology, evolution, and systematics}, vol.~41, pp.
  379--406, 2010.

\bibitem{abc_evolution2}
M.~Mondal, J.~Bertranpetit, and O.~Lao, ``Approximate bayesian computation with
  deep learning supports a third archaic introgression in asia and oceania,''
  \emph{Nature communications}, vol.~10, no.~1, pp. 1--9, 2019.

\bibitem{abc_sysbio}
J.~Liepe, P.~Kirk, S.~Filippi, T.~Toni, C.~P. Barnes, and M.~P. Stumpf, ``A
  framework for parameter estimation and model selection from experimental data
  in systems biology using approximate bayesian computation,'' \emph{Nature
  protocols}, vol.~9, no.~2, pp. 439--456, 2014.

\bibitem{hamra2013markov}
G.~Hamra, R.~MacLehose, and D.~Richardson, ``Markov chain monte carlo: an
  introduction for epidemiologists,'' \emph{International journal of
  epidemiology}, vol.~42, no.~2, pp. 627--634, 2013.

\bibitem{sim_infer_survey}
\BIBentryALTinterwordspacing
K.~Cranmer, J.~Brehmer, and G.~Louppe, ``The frontier of simulation-based
  inference,'' \emph{Proceedings of the National Academy of Sciences}, 2020.
  [Online]. Available:
  \url{https://www.pnas.org/content/early/2020/05/28/1912789117}
\BIBentrySTDinterwordspacing

\bibitem{Warne2020_github}
D.~J. Warne and C.~Drovandi, ``covid19-auto-reg-model (github repository),'' 08
  2020,
  \url{https://github.com/davidwarne/covid19-auto-reg-model/tree/06f25ca5ca567d0795a72ebd411ec2f468cacc6b}.

\bibitem{OG_ABC_IPU}
S.~{Kulkarni}, A.~{Tsyplikhin}, M.~M. {Krell}, and C.~A. {Moritz},
  ``Accelerating simulation-based inference with emerging ai hardware,'' in
  \emph{2020 International Conference on Rebooting Computing (ICRC)}, 2020, pp.
  126--132.

\bibitem{kulkarni2020hardwareaccelerated}
S.~Kulkarni, M.~M. Krell, S.~Nabarro, and C.~A. Moritz, ``Hardware-accelerated
  simulation-based inference of stochastic epidemiology models for covid-19,''
  2020.

\bibitem{tau_leaping}
D.~T. Gillespie, ``Approximate accelerated stochastic simulation of chemically
  reacting systems,'' \emph{The Journal of chemical physics}, vol. 115, no.~4,
  pp. 1716--1733, 2001.

\bibitem{Dong2020}
E.~Dong, H.~Du, and L.~Gardner, \emph{{An interactive web-based dashboard to
  track COVID-19 in real time}}.\hskip 1em plus 0.5em minus 0.4em\relax Lancet
  Publishing Group, may 2020, vol.~20.

\bibitem{ABC_paper}
\BIBentryALTinterwordspacing
M.~Sunnåker, A.~G. Busetto, E.~Numminen, J.~Corander, M.~Foll, and
  C.~Dessimoz, ``Approximate bayesian computation,'' \emph{PLOS Computational
  Biology}, vol.~9, no.~1, pp. 1--10, 01 2013. [Online]. Available:
  \url{https://doi.org/10.1371/journal.pcbi.1002803}
\BIBentrySTDinterwordspacing

\bibitem{Drovandi2011}
\BIBentryALTinterwordspacing
C.~C. Drovandi and A.~N. Pettitt, ``{Estimation of Parameters for Macroparasite
  Population Evolution Using Approximate Bayesian Computation},''
  \emph{Biometrics}, vol.~67, no.~1, pp. 225--233, mar 2011. [Online].
  Available: \url{http://doi.wiley.com/10.1111/j.1541-0420.2010.01410.x}
\BIBentrySTDinterwordspacing

\bibitem{MHAlgo}
\BIBentryALTinterwordspacing
W.~K. Hastings, ``{Monte Carlo sampling methods using Markov chains and their
  applications},'' \emph{Biometrika}, vol.~57, no.~1, pp. 97--109, 04 1970.
  [Online]. Available: \url{https://doi.org/10.1093/biomet/57.1.97}
\BIBentrySTDinterwordspacing

\bibitem{vae}
D.~P. Kingma and M.~Welling, ``Auto-encoding variational bayes,'' 2014.

\bibitem{CMBSim}
M.~Liguori, S.~Matarrese, and L.~Moscardini, ``High-resolution simulations of
  non-gaussian cosmic microwave background maps in spherical coordinates,''
  \emph{The Astrophysical Journal}, vol. 597, no.~1, p.~57, 2003.

\end{thebibliography}
%



%

\begin{IEEEbiography}[{\includegraphics[width=1in,height=1.25in,clip,keepaspectratio]{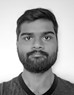}}]{Sourabh Kulkarni}
is currently a research assistant in Nanofabrics and Nanoscale Cognitive Architectures Lab in University of Massachusetts, Amherst. He received the B.Tech. degree in electronics and communication engineering from the Rajarambapu Institute of Technology (RIT), Maharashtra, India, and M.S.E.C.E. degree from University of Massachusetts Amherst in 2015 and 2017, respectively.  He is currently working toward the Ph.D. degree in electrical and computer engineering at University of Massachusetts, Amherst. His research interests include simulation-based inference, hardware achitectures for AI, deep probabilistic programming, and ML systems for life science.
\end{IEEEbiography}


\begin{IEEEbiography}[{\includegraphics[width=1in,height=1.25in,clip,keepaspectratio]{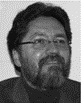}}]{Csaba Andras Moritz}
received the Ph.D. degree in computer systems from the Royal Institute of Technology, Stockholm, Sweden, in 1998. From 1997 to 2000, he was a Research Scientist with Laboratory for Computer Science, the Massachusetts Institute of Technology (MIT), Cambridge. He has consulted for several technology companies in Scandinavia and held industrial positions ranging from CEO, to CTO, and to founder. His company, BlueRISC Inc, develops security microprocessors, hardware-assisted security and system assurance solutions for anti-tamper and cyber defense. He is currently a Professor with the Department of Electrical and Computer Engineering, University of Massachusetts, Amherst. His work spans new models of computing and associated nanoscale integrated circuits. His other interests are in cognitive cyber-security (bluerisc.com) and privacy (eprivo.com).
\end{IEEEbiography}




\end{document}